\documentclass[twocolumn,twocolappendix]{aastex63}
\usepackage[utf8]{inputenc}
\usepackage{enumitem}
\usepackage{graphicx}
\usepackage{hyperref}
\usepackage{mathtools}
\usepackage{comment}
\usepackage{amsmath}
\usepackage{ amssymb }
\usepackage{lipsum} 
\usepackage{float}
\usepackage{commath}
\usepackage{longtable}
\usepackage{cleveref}
\usepackage[caption=false]{subfig}
\usepackage{placeins}

\begin{document}
\title{Statistics of the Chemical Composition of Solar Analog Stars and Links to Planet Formation} 

\author[0000-0001-8042-5794]{Jacob Nibauer}\email{jnibauer@sas.upenn.edu}
\affiliation{Center for Particle Cosmology, Department of Physics and Astronomy,\\ University of Pennsylvania, Philadelphia, PA 19104, USA}

\author{Eric J. Baxter} 
\affiliation{Institute for Astronomy, University of Hawai’i, 2680 Woodlawn Drive, Honolulu, HI 96822, USA}

\author{Bhuvnesh Jain}
\affiliation{Center for Particle Cosmology, Department of Physics and Astronomy,\\ University of Pennsylvania, Philadelphia, PA 19104, USA}

\author[0000-0002-4284-8638]{Jennifer L. van Saders} 
\affiliation{Institute for Astronomy, University of Hawai’i, 2680 Woodlawn Drive, Honolulu, HI 96822, USA}

\author[0000-0002-1691-8217]{Rachael L. Beaton}
\altaffiliation{Hubble Fellow}
\altaffiliation{Carnegie-Princeton Fellow}
\affiliation{Department of Astrophysical Sciences, 4 Ivy Lane, Princeton University, Princeton, NJ 08544, USA}
\affiliation{The Observatories of the Carnegie Institution for Science, 813 Santa Barbara St., Pasadena, CA~91101}
\

\author{Johanna K. Teske} 
\affiliation{Earth and Planets Laboratory
Carnegie Institution of Washington, 5241 Broad Branch Road, N.W., Washington, DC 20015 }

\begin{abstract}
The Sun has been found to be depleted in refractory (rock-forming) elements relative to nearby solar analogs, suggesting a potential indicator of planet formation.
Given the small amplitude of the depletion, previous analyses have primarily relied on high signal-to-noise stellar spectra and a strictly differential approach to determine elemental abundances. 
We present an alternative, likelihood-based approach that can be applied to much larger samples of stars with lower precision abundance determinations. We utilize measurements of about 1700 solar analogs from the Apache Point Observatory Galactic Evolution Experiment (APOGEE-2) and the stellar parameter and chemical abundance pipeline (ASPCAP DR16).  
By developing a hierarchical mixture model for the data, we place constraints on the statistical properties of the elemental abundances, including correlations with condensation temperature and the fraction of stars with refractory element depletions.  We find evidence for two distinct populations: a depleted population of stars that makes up  the majority of solar analogs including the Sun, and a not-depleted population that makes up between $\sim 10-30\% $ of our sample. 
We find correlations with condensation temperature generally in agreement with higher precision surveys of a smaller sample of stars. Such trends, if robustly linked to the formation of planetary systems, provide a means to connect stellar chemical abundance patterns to planetary systems over large samples of Milky Way stars.
\vspace{1cm}
\end{abstract} 

\section{Introduction}

High precision spectroscopy reveals that the Sun may have an unusual elemental composition compared to the majority of nearby solar analogs. In particular, the Sun is found to have a lower than average ratio of refractory to volatile elements, as first discovered by \citet{Mel_ndez_2009}. In that work, the chemical composition of 11 stars nearly identical to the Sun was determined using a line-by-line, strictly differential spectral analysis that allowed for $\sim 0.01~\rm{dex}$ precision on individual abdundance measurements. The depletion of refractory elements found in the Sun was found to correlate with the 50\% condensation temperature ($T_c$)---the temperature at which 50\% of an element condenses from gaseous to solid phase---of each element under protoplanetary disk conditions \citep{2003ApJ...591.1220L}. The $T_c$ trend has spurred extensive debate, with some studies able to verify the anomalous solar composition while others are unable to detect similar depletion patterns across various samples of stars \citep{refId0,2011ApJ...740...76R,Gonzalez_2010,2012A&A...543A..29M,refId02,2011ApJ...732...55S,2014MNRAS.442L..51L,2014ApJ...790L..25T,2015ApJ...815....5S,2015A&A...579A..52N,2016ApJ...819...19T,2018ApJ...865...68B,2020A&A...640A..81N,2020MNRAS.495.3961L}. 

Notably, \citet{refId0} found evidence of a bimodality in the abundance vs. $T_c$ slope distribution at super-solar metallicities, with some stars displaying solar-like depletions while others appear comparatively enriched in refractory elements relative to the more volatile elements. Various mechanisms for the depletion trends have been proposed, the most intriguing of which suggests that missing refractory material could be locked up in rocky planets.
This explanation was initially proposed by \cite{Mel_ndez_2009} and later supported by \cite{2010ApJ...724...92C}, who demonstrated that the Sun's refractory deficit corresponds to roughly four Earth masses of terrestrial material. Other explanations include the possibility of refractory enrichment of stars by planet engulfment  \citep{2011ApJ...740...76R,2015A&A...582L...6S,2018ApJ...854..138O,2020MNRAS.491.2391C}, or interactions within the parent molecular cloud leading to the observed refractory depletion \citep{2010Ap&SS.328..185G}. More recently, \citet{2020MNRAS.493.5079B} utilized evolutionary models for protoplanetary disks surrounding young stars to show that the depletion trend might emerge from a gap in the disk created by a forming giant planet. Under this hypothesis, the gap forms a pressure trap that impedes the accretion of refractory material onto the host star, thereby resulting in the observed deficiency of refractory material. 
Galactic chemical evolution and consequently stellar age have also been proposed as a possible explanation for these trends \citep{2014A&A...564L..15A,2016A&A...593A.125S}. In a recent work, however, \citet{2018ApJ...865...68B} recovers a solar depletion trend after correcting for Galactic chemical evolution effects in a sample of 79 Sun-like stars. 

The standard approach for quantifying trends in the abundance vs. $T_c$ plane involves performing linear fits to the abundance data for each star to derive estimates of the slope and intercept of the abundance trend, which are then compared between stars. This approach requires the determination of elemental abundances to high precision, since the magnitude of the refractory deficit is of order $0.05~\rm{dex}$ \citep{Mel_ndez_2009}. Additionally, in order to reduce systematic uncertainties stemming from Galactic chemical evolution and potentially inaccurate stellar models, most studies focus entirely on solar-like stars, or on stars in binary systems, thus severely limiting sample sizes. Consequently, while previous works have obtained high precision, they have been limited to smaller samples with $\sim$ 20--100 stars. 

Several authors have suggested the importance of increasing sample sizes to build up the statistics for the presence or absence of depletion trends across many stars \citep[i.e.,][]{Mel_ndez_2009,2014A&A...561A...7R,2015ApJ...815....5S}. While elemental abundance measurements from high-resolution spectroscopy are available for hundreds of thousands of stars
from surveys such as the Apache Point Galactic Chemical Evolution Experiment  \citep[APOGEE;][]{2017AJ....154...94M} or the Galactic Archaeology with HERMES survey \citep[GALAH;][]{galah_overview}, 
the precision of these abundance measurements are generally below the threshold required to detect subtle depletion trends for individual stars using the methods previously described.

In this work, we attempt to measure depletion trends across many stars using a novel statistical approach. We utilize abundance data from the sixteenth data release of the Sloan Digital Sky Survey, primarily relying on the second iteration of the APOGEE survey \citep[APOGEE-2;][]{2020ApJS..249....3A,2020arXiv200705537J}. While the uncertainties on elemental abundance data from APOGEE-2 are generally higher than in the previously mentioned studies, we combine measurements across stars using a likelihood-based approach to place constraints on the overall distribution of elemental abundances in solar analogs, rather than relying on the precision measurements of individual stars. In addition to applying our analysis to large samples of stars from APOGEE-2, we also apply our analysis to publicly available line-by-line abundance measurements from \citet{2018ApJ...865...68B} for a sample of 79 Sun-like stars, finding roughly consistent results.

We have previously used a similar formalism to detect circumstellar debris disks in data from the {\it Planck} cosmic microwave background survey, constraining parameters such as the fraction of stars with debris disks \citep{2020AJ....159..210N}. The present work employs a similar modeling scheme, enabling us to place constraints on the fraction of stars with elemental depletions.

The paper is organized as follows. In \S\ref{sec: Data} we describe the data from APOGEE-2 and \citet{2018ApJ...865...68B} used in our analysis; in \S\ref{sec: Analysis} we describe our analysis pipeline and modeling approach; in \S\ref{sec: Results} we present our results on data from APOGEE-2 and \citet{2018ApJ...865...68B}, and we conclude in \S\ref{sec: discussion}.

\begin{figure}[h!]
    \includegraphics[width=8.5cm]{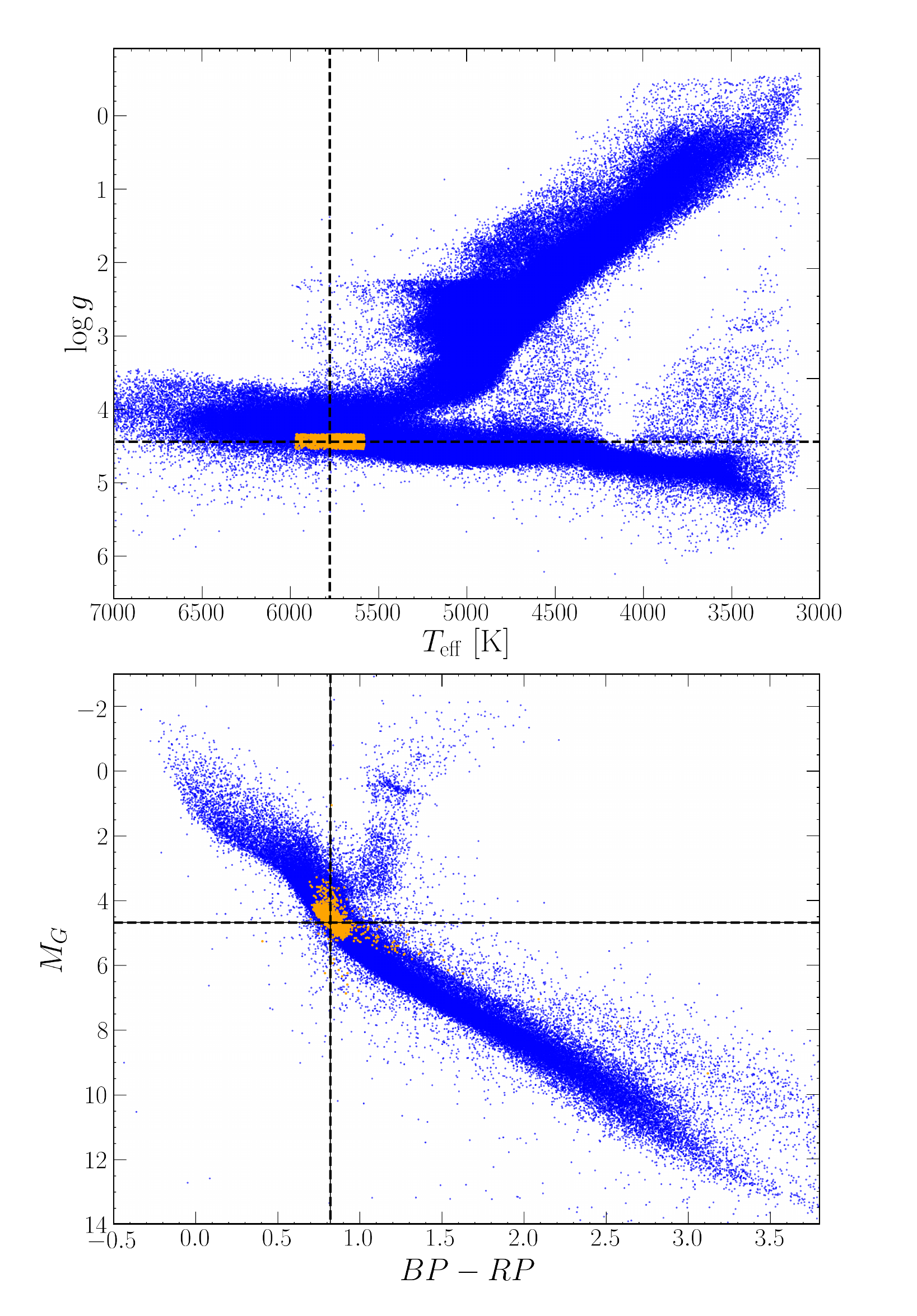}
    \centering
    \caption{Top: Kiel diagram of the APOGEE sample. Orange points indicate the population of Sun-like stars used in our analysis of APOGEE data, as described in \S\ref{sec: star selection}. Horizontal and vertical dashed lines are placed at solar values for reference. Bottom: Color-Magnitude diagram with parallax, color, and magnitude from \textit{Gaia} DR2. Orange points correspond to the same set of stars in the top panel, while blue points represent all APOGEE-2 targeted stars within 350~pc. Horizontal and vertical lines correspond to solar values for $M_G$ and $BP-RP$, respectively.}\label{fig:HR Diagram}
\end{figure}

\section{Data}\label{sec: Data}
\subsection{APOGEE}\label{sec: Data APOGEE}
We utilize data from the sixteenth release of the Sloan Digital Sky Survey (SDSS), primarily relying on the second iteration of APOGEE  \citep[APOGEE-2, hereafter referred to as APOGEE;][]{2017AJ....154...28B,2017AJ....154...94M,2020ApJS..249....3A,2020arXiv200705537J}. APOGEE has produced $H$-band spectra of over 450,000 stars, allowing for the determination of stellar parameters and chemical abundances to trace the history and evolution of the Milky Way. APOGEE includes data taken from both hemispheres, using the Sloan Foundation 2.5-m telescopes from Apache Point Observatory and the Ir\'en\'ee du Pont telescope from Las Campanas Observatory \citep{2006AJ....131.2332G,1973ApOpt..12.1430B}, with targeting procedures described in \citet{2013AJ....146...81Z,2017AJ....154..198Z}. In particular, our study relies on data obtained in Contributed Programs with APOGEE-2S (F.~Santana et al.~in prep.) and in the Bright Time Extension of APOGEE-2N (R.~Beaton et al.~in prep.).

Stellar parameters and chemical abundances are determined by the APOGEE Stellar Parameter and Chemical Abundance Pipeline \citep[ASPCAP;][]{2016AJ....151..144G, 2018AJ....156..125H,2020arXiv200705537J}, based on the \texttt{FERRE} code for spectroscopic analysis \citep{2006ApJ...636..804A}. The ASPCAP pipeline first determines fundamental stellar parameters over the entire spectral range. Then, by adopting the best fit parameters a sequential fit for each elemental abundance is performed, using limited spectral windows matched to the element of interest. This procedure is repeated for each star, with a line list enabling the determination of abundances for up to 26 species \citep[][V.~Smith et al.~in prep.]{Shetrone_2015}. The reliability and precision of the abundance determinations varies widely across stars and elements. We therefore make use of a limited sample in order to minimize the effects of systematic errors (see \S\ref{sec: star selection} and \S\ref{sec: element selection}). Detailed discussions for all elements included in ASPCAP DR16 and further information on the ASPCAP pipeline can be found in \citet{2020arXiv200705537J}. 

\subsubsection{Star Selection}\label{sec: star selection}
We select solar analogs by roughly following the criteria used in \citet{2010MNRAS.407..314G} and \citet{2018ApJ...865...68B}. In particular, we include stars with $\rm{distance}~<350~\rm{pc}$, $\abs{\rm{[Fe/H]}} < 0.1~\rm{dex}$, $\log{g}$ within 0.1~dex of solar, and $T_{\rm{eff}}$ within $195~\rm{K}$ of solar. 
Distances are determined using parallax measurements from \textit{Gaia} DR2 \citep{2018A&A...616A...1G}, while all other stellar parameters are from the calibrated values produced by the ASPCAP pipeline. The APOGEE-2 ASPCAP pipeline provides a series of bitmask flags to indicate stars with potential issues. Accordingly, we select stars with ASPCAPFLAG=0, and make exclusive use of the $\rm{X\_FE}$ tagged columns since these are only populated for the most reliable spectra. The final population of $\sim 1800$ stars is illustrated by the orange points in Fig.~\ref{fig:HR Diagram}, where the dashed black lines depict solar values.  Our selection criteria firmly places our sample on the main sequence and sufficiently far from the sub-giant branch to ensure that we are sampling dwarf-type stars. Over this $\log{g}$-$T_{\rm eff}$ range, a single calibration is used for the stellar parameters \citep{2020arXiv200705537J}.

A small fraction of our selected stars scatter in the direction of high $BP - RP$ and $M_G$ (see orange points in Fig.~\ref{fig:HR Diagram}, bottom panel), likely a result of line-of-sight reddening, which has not been corrected for in this diagram.\footnote{APOGEE uses de-reddened 2MASS photometry to select stars using the Rayleigh-Jeans Color Excess method (RJCE;  \citealt{2013AJ....146...81Z,2011ApJ...739...25M}) and the sample is therefore minimally biased by line-of-sight extinction variations.} We find that removing these stars from our analysis has no significant impact on our results.

\subsubsection{APOGEE Abundances and Element Selection}\label{sec: element selection}
In order to maximize our ability to detect subtle elemental depletion patterns, we exclude the lower $T_c$ (volatile) elements that are not expected to fall along the steepest portion of the depletion trend \citep{Mel_ndez_2009,2010ApJ...724...92C,2018ApJ...865...68B}. The distinction between refractory and volatile elements has been roughly defined at $T_c \sim 900~\rm{K}$, where elements with condensation temperatures greater than this threshold have been found to 
fall along the steepest portion of the abundance vs. $T_c$ trend. \citep{Mel_ndez_2009,refId0,2018ApJ...865...68B}.
 
APOGEE provides chemical abundance measurements for up to 26 chemical species, spanning a wide range of condensation temperatures. Some species have significantly higher precision abundance determinations than others, in part due to the difficulty in measuring elemental compositions when there are few clean spectral features for a given species. To achieve the highest degree of precision possible in our study, we use a subset of high $T_c$ elements with the most reliable abundance determinations. Based on the discussions presented in \citet{2020arXiv200705537J}, Si, Mg, Ni, Ca, and Al are deemed the five most precisely determined elemental abundances with high $T_c$, for dwarf stars in particular. We therefore place constraints on $T_c$ trends with these elements using APOGEE data in \S\ref{sec: APOGEE Results}.

Because we will compare our study to external results, it is important to understand how the APOGEE data is calibrated. 
APOGEE works with stars spanning a wide range of stellar parameters, and has thus tuned their synthetic stellar atmospheres to both the solar spectrum (a G dwarf) and to Arcturus (a K giant) \citep[see e.g.,][Smith et al.~in prep.]{Shetrone_2015}.
Internal uncertainties are calibrated using serendipitous duplicate observations of the same stars that were reduced independently \citep{2020arXiv200705537J,2020arXiv200906777P}.
Stellar parameters are calibrated against external estimates, with $\log g$ calibrated to seismic measurements and $T_{\rm eff}$ calibrated against photometric temperatures \citep[for details see][]{2020arXiv200705537J}. 
The APOGEE abundances are not adjusted after the calibration of the stellar parameters, but are instead calibrated by the determination of a zero-point using stars in the solar-neighborhood (defined as those stars within 500 pc).
This calibration sample uses all stars of all stellar types and not just solar-type stars.
\citet{2020arXiv200705537J} provides the residuals for the solar spectrum reflected off of the asteroid Vesta as a means of assessing the offsets between the APOGEE calibration and calibrations directly to the  solar spectrum. 

The APOGEE calibration methods are motivated by numerous results in the literature suggesting that stars near solar abundance in the solar neighborhood have mean [X/Fe] = 0 (see \citealt{2020arXiv200705537J} and references therein). It should be emphasized that this calibration is simply a zero-point offset applied to each element, meaning that intrinsic scatter in the abundances is unaffected. Additionally, the size of the offset is small for the five elements selected for in this work, with the largest zero-point shift of $0.043~\rm{dex}$ for Al and Ni \citep[see][their table 4]{2020arXiv200705537J}.
The result is that the APOGEE abundances are explicitly calibrated to the solar-neighborhood, which makes the absolute abundance scale ``solar-like''.\footnote{The abundance scale has some dependence on assumptions made in both the construction of synthetic spectra and in spectral fitting; these concerns are discussed at length in \citet{2020arXiv200705537J} for DR16 with references to prior work on ASPCAP.} 

Uncertainties on the APOGEE abundances are also derived statistically, based on a sample of $\sim$15000 stars with duplicate observations that were processed independently and are then separated into samples representative of ranges of stellar type by $T_{\rm eff}$ and $\log g$. 
Lastly, in our study we use the values in the ``named tags'' and, as described in \citet{2020arXiv200705537J}, these tags are only populated if the underlying spectral data and the APOGEE methods applied to those stellar types are trustworthy.

\subsection{Data from Bedell et al. and Spina et al.}\label{sec: Bedell Data}
In addition to our analysis on APOGEE data, we utilize abundance measurements from \citet{2018ApJ...865...68B} and \citet{2018MNRAS.474.2580S} for a sample of 79 Sun-like stars with similar $T_{\rm eff}$, $\log{g}$, and [Fe/H] to our APOGEE sample.  These stars were observed with the high resolution HARPS spectrograph \citep{2003Msngr.114...20M}, and abundances were obtained using the strictly differential line-by-line technique described in \citet{2014ApJ...795...23B}. This enables high signal-to-noise abundance determinations for more than 20 elements, allowing us to place constraints on subtle depletion trends in the data. For a more detailed description of these data, we refer readers to \citet{2018ApJ...865...68B} and \citet{2018MNRAS.474.2580S}.

\section{Analysis}\label{sec: Analysis}

\subsection{Modeling}\label{sec: modeling}

We now develop a model for the stellar elemental abundance measurements discussed above.  Our primary goal in modeling the abundances is to determine whether there is evidence for two populations of stars: one which is depleted in high $T_c$ refractory elements relative to the more volatile elements, and another which is comparably less deficient in high $T_c$ elements relative the volatiles.

Motivated both by simplicity and by several previous works (e.g.,  \citealt{Mel_ndez_2009,refId0,2018ApJ...865...68B}), we assume that the  elemental abundance value for an individual star is a linear function of the element condensation temperature, $T_c$, over the narrow range of $T_c$ values considered.  This assumption is made for both the depleted and non-depleted populations.  In other words, we assume
\begin{equation}\label{eq: linear_abundance}
d^{\rm obs}_j = d_j^{\rm true} + N_j =  m T_{c,j}+b+N_j
\end{equation} 
where $d^{\rm obs}_j$ is the abundance measurement for the $j^{\rm th}$ element, $d_j^{\rm true}$ is the true abundance value for that element, $N_j$ represents a noise term to be discussed later, and $m$ and $b$ are parameters of the model. 
The model in Eq.~\ref{eq: linear_abundance} is likely to be a good approximation as long as the set of elements considered is strictly refractory and therefore does not span a very large range in $T_c$ \citep{2018ApJ...865...68B,2018ApJ...854..138O, 2018ApJ...853...83H}. 

Since we are attempting to model the abundance distribution as the sum of two populations of stars, we adopt a mixture model for describing the measurements from all stars.  We refer to the fraction of stars with depleted abundances (i.e. with lower abundances of refractory elements relative to the more volatile elements) as $f$, so the fraction of stars without depletion is $1-f$.  The values of $m$ and $b$ for each star are assumed to be drawn from two distributions corresponding to the depleted and not-depleted model components.  For simplicity, we assume that the distributions of $m$ and $b$ values for both components are Gaussians:
\begin{equation}\label{eq:m b distributions}
    P^{\rm D}(\{\rm m,b\}) = \mathcal{N}(\{m,b\} | \mu = \mu^{\rm D}_{\{m, b\}}, \sigma = \sigma^{\rm D}_{\{m,b\}}),
\end{equation}
where $\rm D$ stands for depleted, and we use the notation $\{ A, B\}$ to represent a choice of either $A$ or $B$. We note that the formalism for not-depleted (ND) stars appears exactly the same as for D stars, up to a superscript (D is replaced with ND for not-depleted stars in Eq.~\ref{eq:m b distributions}).

Once $m$ and $b$ are drawn for a star, all that remains to connect the model to the data is a description of the noise term in Eq.~\ref{eq: linear_abundance}.  We refer to the observed abundance measurement for the $i^{\rm th}$ star and $j^{\rm th}$ element as $d^{\rm obs}_{ij}$ (i.e. [X/Fe] measured in dex).  We write the noise for the $i^{\rm th}$ star and $j^{\rm th}$ element as the sum of two noise terms:
\begin{equation}\label{eq: linear relation}
    N_{ij} =  N_{{\rm intr},j} + N_{{\rm inst},ij}.
\end{equation}
where $N_{{\rm intr},j}$ and $N_{{\rm inst},ij}$ represent intrinsic scatter and instrumental scatter in the abundance measurements, respectively.  The intrinsic scatter for each element is assumed to be constant across all stars, while the instrumental scatter varies element-to-element and star-to-star.  For the intrinsic scatter, we treat the dispersion as a free parameter of the model denoted by $\sigma_{\mathrm{intr},j}$, which can be different for both depleted and not-depleted stars (denoted by $\sigma_{\mathrm{intr},j}^{\rm D}$ and $\sigma_{\mathrm{intr},j}^{\rm ND}$, respectively).  For the instrumental scatter, we simply adopt the uncertainties reported by APOGEE (or \citealt{2018ApJ...865...68B} in \S\ref{sec: Bedell Results}), $\sigma_{\mathrm{inst},ij}$.

\begin{figure}[tp!]
    \includegraphics[width=8.5cm]{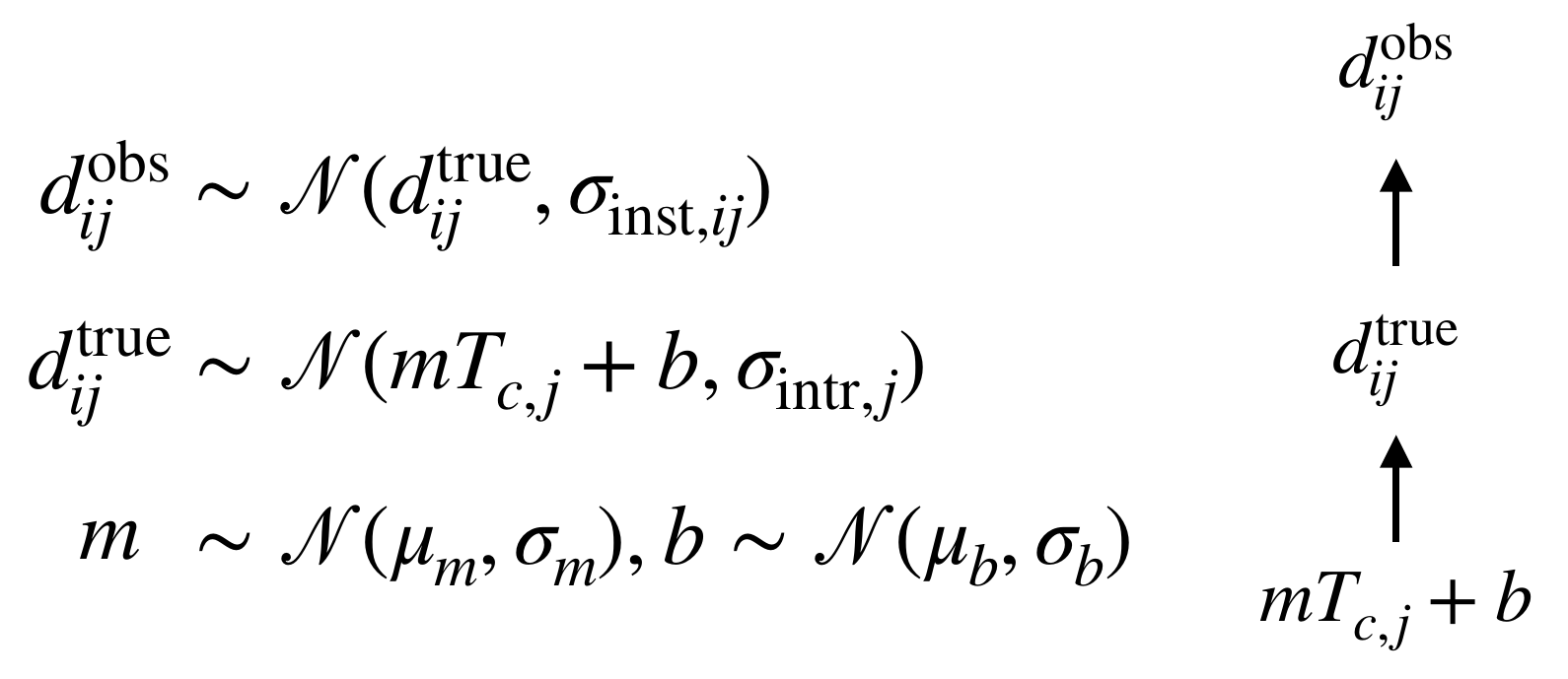}
    \centering
    \caption{Directed acyclic graph representing the hierarchical mixture model described in \S\ref{sec: modeling}. Note that the architecture of the model is the same for both mixture components (D vs. ND), with the only difference being separate parameters for all $\mu_{\{m,b\}}$, $\sigma_{\{m,b\}}$, and $\sigma_{{\rm intr,}j}$.}\label{fig: DAG Visual}
\end{figure}
A directed acyclic graph representing either the depleted or not-depleted model components of the hierarchical model is shown in Fig.~\ref{fig: DAG Visual}.  This model allows us to write the likelihood for the data as
\begin{multline}\label{eq: marginalize d_true}
    P^{\rm D}\left(d^{\rm obs}_{ij}| \theta^{\rm D} \right) = \\
    \int d(d_{ij}^{\rm true})\,dm\,db \,
    P\left(d^{\rm obs}_{ij}|d_{ij}^{\rm true}, \sigma_{\mathrm{inst},ij} \right) P\left(d_{ij}^{\rm true}|m,b,\sigma_{\mathrm{intr},j}^{\rm D} \right) \\ \times P^{ D }(m|\theta^{D})P^{\rm D}(b|\theta^{\rm D}),
\end{multline}
where we have marginalized over the (unobservable) values of $d^{\rm true}$, and $\theta^{\rm D}$ represents parameters describing the distributions (i.e. $\mu^{\rm D}_{\{m, b \}}$ and $\sigma^{\rm D}_{\{m, b\}}$). We note again that Eq.~\ref{eq: marginalize d_true} is the same for not-depleted stars, up to the D superscript (D is replaced with ND for the not-depleted case).

Assuming the measurement of each element is statistically independent, we may write the likelihood for the $i^{\rm th}$ star as a product over the element index $j$:
\begin{multline}
    \mathcal{L}_i(d^{\rm obs}_{i}|\theta) = f \prod_j P^{\rm D}(d^{\rm obs}_{ij}|\theta^{\rm D}) + \\ (1-f) \prod_j P^{\rm ND}(d^{\rm obs}_{ij}|\theta^{\rm ND}),
\end{multline}
where $\theta$ represents the complete set of model parameters, and $d^{\rm obs}_{i}$ represents the abundance data for the $i^{\rm th}$ star. Next, by assuming the measurement of each star is statistically independent, we may write the likelihood for the data as a product over all stars:
\begin{equation}\label{eq:final_likelihood}
    \mathcal{L}\left(\{d_{ij}^{\rm obs} \}|\theta\right) = \prod_i \mathcal{L}_i(d^{\rm obs}_i|\theta).
\end{equation}
We adopt uniform priors on the parameters in our model, so the posterior on these parameters is simply proportional to the likelihood in Eq.~\ref{eq:final_likelihood}.

\subsection{Outlier Rejection}\label{sec: outlier rejection}

We assume that the distributions of $m$ and $b$ are Gaussian.  In order to ensure that the model fit is not being driven by departures from our  Gaussian assumption, particularly in the tails of the abundance distributions, we employ outlier rejection. In particular, we calculate the standard deviation of the abundance measurements for each element, $\sigma \left(\rm{[X/Fe]} \right)$, after the selections described in \S\ref{sec: star selection} and \S\ref{sec: element selection}. We adopt a fiducial outlier rejection of $2\sigma$, such that any star with $\abs{\rm{[X/Fe]}-\langle\rm{[X/Fe]}\rangle} > 2\sigma \left(\rm{[X/Fe]} \right)$ is removed from the analysis. Since the expected depletion trend is of the order $\sim 0.05~\rm{dex}$, the signal of interest should not be driven by the tails of abundance distributions regardless. Outlier rejection removes roughly 200 stars from our Sun-like sample, with 1556 remaining. We note that on simulated data, we recover the correct input parameters with $2\sigma$ outlier rejection.

We explore variations on the $2\sigma$ outlier rejection in \S\ref{sec: Analysis of Simulated Data} for simulated data and in Appendix~\ref{sec: variations on outlier rejection} for actual APOGEE data. In both cases, we find that model constraints are robust to various rejection choices overall.

\subsection{Analysis of Simulated Data}\label{sec: Analysis of Simulated Data}

We test our analysis pipelines and evaluate the extent to which we are able to recover input parameters with realistic levels of noise using simulated data.  We generate mock data by drawing from the distributions described in \S\ref{sec: modeling}, using the real APOGEE uncertainty estimates. Mock abundances are simulated for the five elements used in the data analysis.
We assume $\sim 2000$ stars, meant to roughly match the APOGEE sample discussed in \S\ref{sec: star selection}. 

The true parameter values describing the distributions of $m$ and $b$ are chosen based on the distribution of abundance vs. $T_c$ slopes provided by \citet{2018ApJ...865...68B} for a Sun-like sample. The intrinsic dispersion for each mock element (given by $\sigma_{{\rm intr},j}$) is chosen to fall between 0.02 and 0.05~dex for both depleted and not-depleted stars, since this is roughly the range of standard deviations for a typical chemical abundance distribution from our APOGEE solar analog sample. Our fiducial parameter choice is $f = 0.5$,  based on the large population of stars near solar abundances in \citet{refId0,2018ApJ...865...68B}.

Results on simulated data are presented in Appendix~\ref{app: analysis of sim data}.
We find that for five elements with uncertainties sampled directly from APOGEE data, all input parameters are recovered without bias. Additionally, when simulating data with only a single population of stars and APOGEE abundance uncertainties, the analysis correctly recovers $f=0$. 

We test our sensitivity to the assumptions of Gaussian distributions in Appendix~\ref{sec: Departures from Gaussian Assumption}, and find our analysis to be generally robust to small departures from Gaussinity. 

\section{Results}\label{sec: Results}
\begin{figure*}[ht!]
    \includegraphics[width=16.5cm]{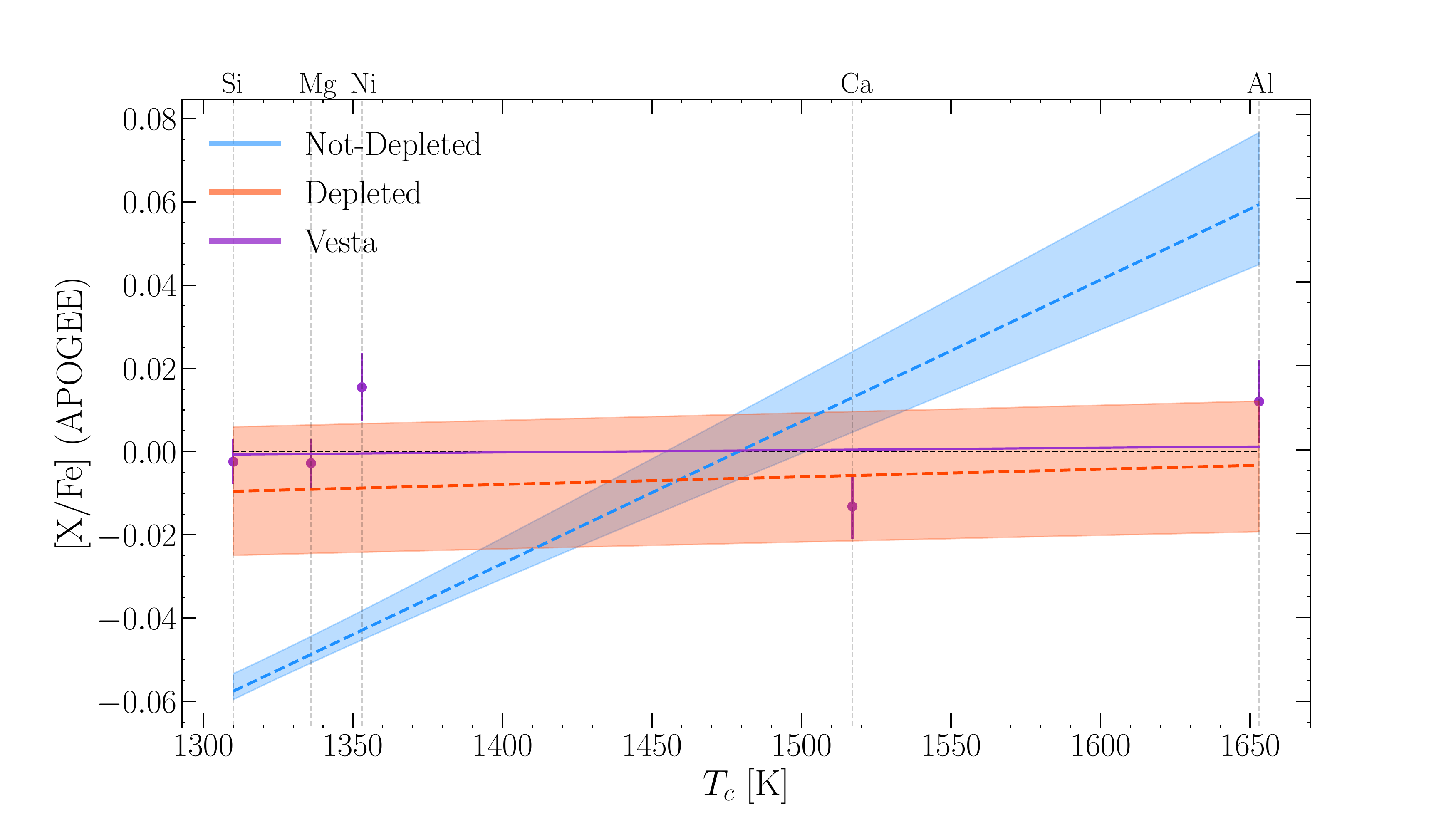}
    \centering
    \caption{Model constraints (68\% CL) from APOGEE data for the elements Si, Mg, Ni, Ca, Al. The [X/Fe]-$T_c$ trend is illustrated by plotting a realization of the model at each step in the MCMC parameter chain. Blue and red correspond to not-depleted and depleted, respectively. Dashed vertical lines are placed at the condensation temperature of each element, while the colored dashed lines depict the best fit mean trend for D and ND stars (i.e. $\mu_m T_c + \mu_b$). The APOGEE solar reference spectrum, measured in reflected sunlight from the asteroid Vesta, is shown in purple.}\label{fig:main result}
\end{figure*}

\subsection{Abundances from  APOGEE}\label{sec: APOGEE Results}
We now present the results of applying the statistical methodology developed in \S\ref{sec: Analysis} to APOGEE data for the selection of stars described in \S\ref{sec: Data}, using the five elements Si, Mg, Ni, Ca, and Al. 

In Fig.~\ref{fig:main result}, we illustrate our constraints on the elemental abundance of solar analogs as a function of element condensation temperature, $T_c$. Because APOGEE calibrates to solar neighborhood stars (see \S\ref{sec: Data APOGEE}), we also plot abundances from the APOGEE solar reference spectrum (purple) which is measured in reflected sunlight from the asteroid Vesta \citep{2020arXiv200705537J}. A linear fit to these points is illustrated by the purple line. We note that the solar reference spectrum is roughly consistent with an abundance vs. $T_c$ slope and intercept of zero (horizontal dashed line in Fig.~\ref{fig:main result}), indicating that the APOGEE solar neighborhood calibration is not unusually different from calibrations relative to the Sun.\footnote{A more direct comparison to the Sun can be achieved by measuring the raw APOGEE elemental abundances relative to those from the asteroid Vesta. Given the small offsets between the solar neighborhood calibration and the solar reference abundances, however, we postpone this to a future work.} In \S\ref{sec: caveats}, we discuss the potential impact of the APOGEE ASPCAP calibrations on our results and comparisons to other studies. 

We find the mean slope of the abundance vs. $T_c$ relation for not-depleted (ND) and depleted (D) stars to be significantly different, with
\begin{equation}\label{eq: apogee slope constraints}
    \mu_{m} = \begin{dcases*} 
         3.47 \pm ^{0.269}_{0.358}\times 10^{-4}~\rm{dex~K^{-1}}  \ \rm{for \ ND,} \\
        1.65 \pm ^{0.448}_{0.307}\times 10^{-5}~\rm{dex~K^{-1}} \ \rm{for \ D} \\
       \end{dcases*}
    \end{equation}\label{eq: mu_m constraints} 
at 68\% confidence. We find that the D band in Fig.~\ref{fig:main result} is not in fact below the ND band for the entire range of $T_c$. While we are primarily interested in the slope of the abundance vs. Tc relation---from which the ND population is refractory enriched relative to the more volatile elements---the absolute abundances of both populations are potentially complicated by ASCAP calibrations, as discussed in \S\ref{sec: Data APOGEE} and \S\ref{sec: discussion}. Our constraints place an abundance vs. $T_c$ slope of zero (i.e., the Sun's pattern) to reside at roughly 1.2$\sigma$ from the D mean, towards the more refractory depleted tail of the slope distribution. We note that due to degeneracies between $\mu_m$ and $\mu_b$, the marginalized uncertainty on $\mu_m$ does not capture the full picture.  
Instead, it is more instructive to evaluate the ND and D abundances as a function of $T_c$, as shown in Fig.~\ref{fig:main result}. Additionally, in Appendix~\ref{app:posterior on model parameters} we provide contour plots of the model posterior.
\begin{figure}
    \includegraphics[width=8.5cm]{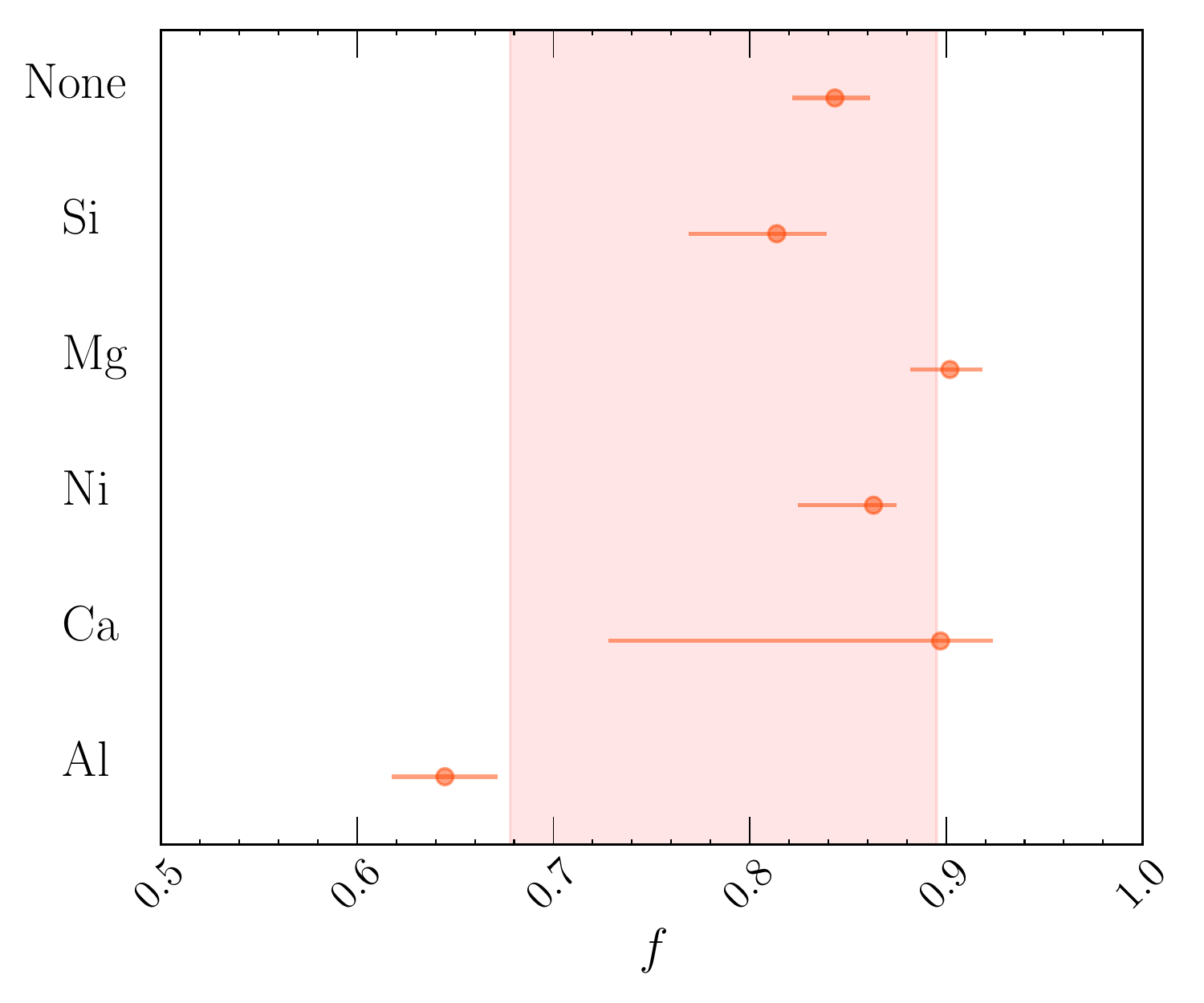}
    \centering
    \caption{Constraints on the fraction ($f$) of stars in the depleted population. In order to evaluate the robustness of our constraint on $f$, we remove each element in turn and repeat the analysis. The labels on the y-axis specify which element is removed for each constraint, with ``None" corresponding to the case for which none of the five elements are removed. The red band corresponds to our estimate of the combined statistical and systematic error (68\% CL).}\label{fig: f constraint}
\end{figure}

We find evidence of two populations of stars -- one with near zero (solar) abundance vs. $T_c$ slopes (D), and another with significantly steeper slopes (ND) indicating a higher ratio of refractory to volatile (i.e. high $T_c$ to low $T_c$) abundances compared to the Sun. In Fig.~\ref{fig: f constraint}, we plot constraints on the fraction ($f$) of stars drawn from the depleted model component.  When incorporating only statistical uncertainty, we find $f = 0.84 \pm^{0.01}_{0.03}$ at 68\% confidence. In order to evaluate the robustness of our fit and provide a systematic error bar on $f$, we repeat the analysis multiple times by removing each element in turn. The labels on the y-axis of Fig.~\ref{fig: f constraint} specify the element removed for each constraint, with ``None" corresponding to the case of no elements removed (i.e. Si, Mg, Ni, Ca, Al all included). In order to derive a systematic error bar from the six constraints illustrated in Fig.~\ref{fig: f constraint}, we add the probability density functions on $f$ from the MCMC parameters chains across all six trials, and determine the 68\% confidence interval on the resulting distribution. This assumes that the true result could be any of the six trials illustrated in Fig.~\ref{fig: f constraint}. Our constraint on the fraction of stars drawn from the depleted distribution is then
\begin{equation}\label{eq: f constraint}
    f = 0.84 \pm ^{0.05}_{0.17}
\end{equation}
 at $68\%$ confidence. 

The magnitude of the mean abundance vs. $T_c$ slopes are generally within the expected range of previous observations, with most studies reporting [X/Fe] vs. $T_c$ slopes between
$\sim -1\times10^{-4} \ \rm{and} \ \sim 3\times10^{-4}$ for a generally wider range of condensation temperatures \citep{Mel_ndez_2009,refId0,2014A&A...564L..15A,2015A&A...579A..52N,2018ApJ...865...68B}. A rough comparison of the mean abundance vs. $T_c$ slope constraints in Eq.~\ref{eq: apogee slope constraints} can be made with \citet{refId0}, who applied a Kolmogorov-Smirnov test to the abundance vs. $T_c$ slopes of $\sim 60$ super-solar metallicity stars, finding evidence of a bimodality at $2.7\sigma$. Their test assumes  two Gaussian distributions centered on $m = -0.5 \times 10^{-4}$ and $0.9 \times 10^{-4}~\rm{dex~K^{-1}}$, respectively. We perform a more detailed comparison with previous works in \S\ref{sec: Tests for Consistency, APOGEE}.

 We emphasize that the parameter $f$ describes the fraction of stars drawn from the depleted population.  Several previous authors have instead considered the fraction of stars which have a lower refractory to volatile element ratio compared to the Sun (i.e. stars with  [X/Fe] vs. $T_c$ slopes $\leq 0$), finding that $\sim 20\%$ of solar analogs are more depleted than the sun \citep{refId0,Gonzalez_2010,2011ApJ...732...55S, 2018ApJ...865...68B}.
 Our constraints on the depleted fraction $f$ suggest that the majority of solar analogs have an abundance vs. $T_c$ relation similar to the Sun. The two findings are consistent: in particular, by simulating data with underlying parameters governed by our model constraints, we find that less than 15\% of the full sample of stars have abundance vs. $T_c$ trends as refractory depleted as the Sun. While previous studies have interpreted this result as the Sun being unusual in its elemental composition relative to the majority of solar analogs (e.g. \citealt{Mel_ndez_2009}), our constraints place the Sun close to the middle of a distribution of stars with similar abundance vs. $T_c$ relations (the depleted population).  We defer a more detailed discussion of our interpretation of these results to \S\ref{sec: discussion}.

We perform a detailed analysis of the goodness of fit of our model in Appendix~\ref{app: goodness of fit}, where we employ posterior predictive methods. We find that for the set of elements presented in this section, the model provides a good fit to the data. We also note that the preference for a bimodal model (i.e. $f \neq 0$) is preferred at high significance on APOGEE data, compared to the model with $f = 0$ and all other parameters free.

\subsection{Comparison to Bedell et al. 2018}\label{sec: Tests for Consistency, APOGEE}
We now compare our results derived from APOGEE data to high-precision measurements from \citet{2018ApJ...865...68B}. 

\subsubsection{Distribution of Abundance vs. $T_c$ Slopes}\label{sec: distribution of abundance vs. Tc slopes}
We generate elemental abundance data with underlying parameters governed by our model constraints on APOGEE in \S\ref{sec: APOGEE Results}. Since we wish to compare the resulting abundance vs. $T_c$ trends to the literature \citep[i.e.][]{2018ApJ...865...68B}, we adopt a scatter of $\sim 0.02 ~\rm{dex}$ on all simulated abundances, propagated through to the distribution of abundance vs. $T_c$ slopes. This scatter is meant to emulate the uncertainty in abundance determinations from generally higher precision surveys, such as \citet{Mel_ndez_2009} and \citet{2018ApJ...865...68B}, thereby enabling a more direct comparison.

\begin{figure}[tp!]
    \includegraphics[width=9cm]{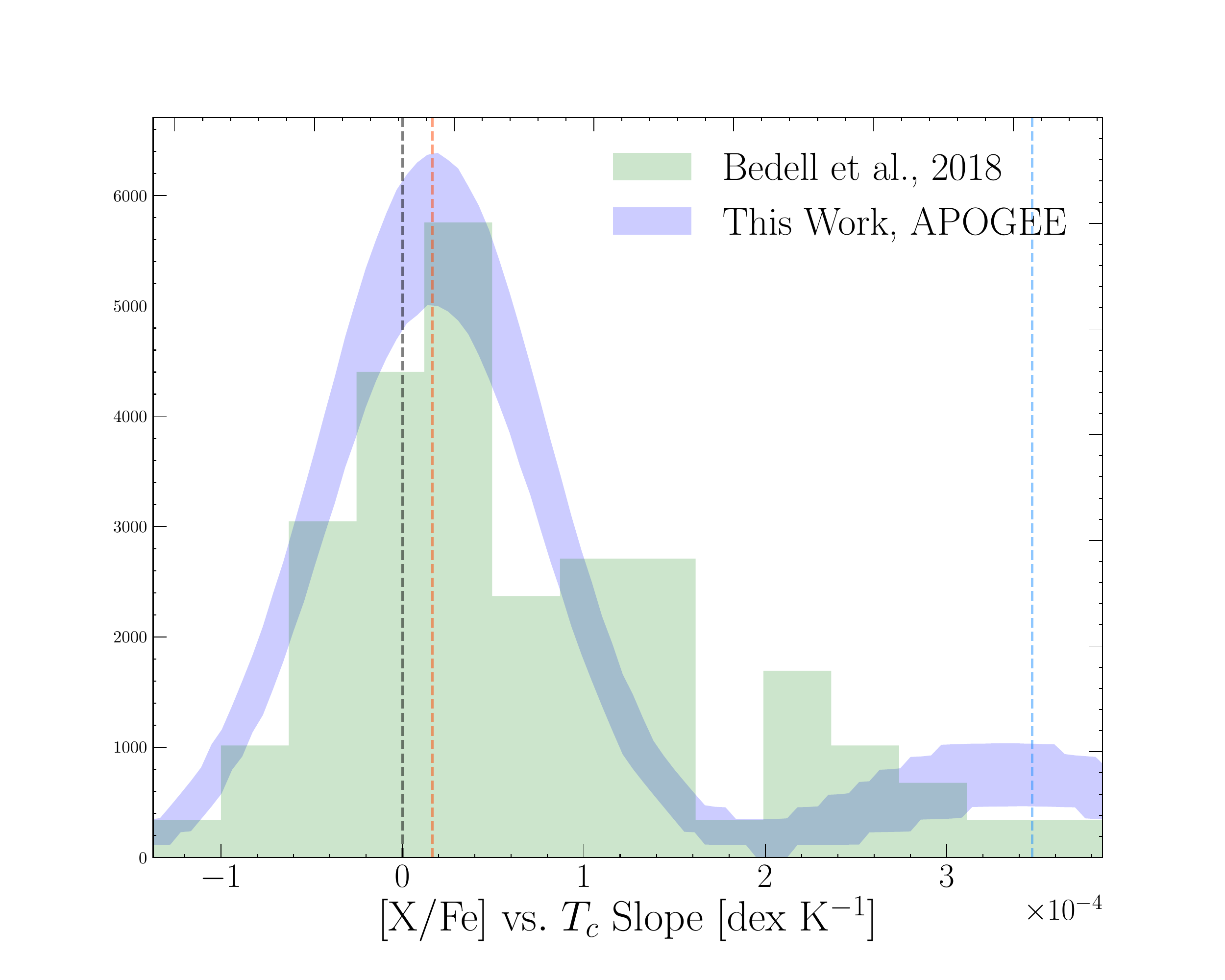}
    \centering
    \caption{Distribution of [X/Fe] vs. $T_c$ slopes from model constraints on APOGEE data (blue). The width of the blue band as a function of [X/Fe] vs. $T_c$ slope is derived from the 68\% confidence interval on the simulated APOGEE data, which has an added abundance uncertainty of $\sim 0.2~\rm{dex}$ meant to emulate errors in other works (see \S\ref{sec: Tests for Consistency, APOGEE}). Blue and red vertical dashed lines are placed at the mean slope values ($\mu_m$) for ND and D stars, respectively. The vertical dashed black line is placed at zero, representing the solar abundance trend. We also include the distribution of slopes from \citet{2018ApJ...865...68B} in green, derived from a linear fit to the abundance data (vs. $T_c$) for a sample of 79 stars and 21 high $T_c$ elements. Both distributions (APOGEE and \citealt{2018ApJ...865...68B}) are normalized to have equal areas, such that the y-axis can be interpreted as a probability density. }\label{fig: Apogee vs. Bedell Slopes}
\end{figure}

The resulting distribution of abundance vs. $T_c$ slopes is illustrated by the blue band in Fig.~\ref{fig: Apogee vs. Bedell Slopes}, where we also include the distribution of $T_c > 1300~\rm{K}$ slopes from \citet{2018ApJ...865...68B} for a smaller sample of 79 Sun-like stars with similar $T_{\rm eff}$, $\log{g}$, and metallicity to our APOGEE sample (green).  Note that the \citet{2018ApJ...865...68B} distribution is derived by performing a linear fit to the abundance data for each star, as is common in the literature. We limit the \citet{2018ApJ...865...68B} abundance data to the 21 elements with $T_c > 1300~\rm{K}$, since this enables a more direct comparison to our constraints on APOGEE (with $T_{c,\rm{min}}\sim 1300~\rm{K}$). Additionally, \citet{2018ApJ...865...68B} finds that including moderately volatile elements with $T_c$ between $900~\rm{K}$ and $\sim 1300~\rm{K}$ can bias the abundance vs. $T_c$ slope.

We find good agreement between 
constraints from our APOGEE sample discussed in \S\ref{sec: APOGEE Results} and results from \citet{2018ApJ...865...68B}, given that the \citet{2018ApJ...865...68B} data includes generally higher precision abundance measurements of many more elements determined from far fewer stars. Furthermore, we find that when adopting a minimal abundance uncertainty of $\sim 0.02 ~\rm{dex}$ and simulating data with model parameters governed by our APOGEE constraints, the fraction of stars with [X/Fe] vs. $T_c$ slopes less than zero (i.e. stars which are {\it more} refractory depleted than the sun relative to the lower $T_c$ elements) is roughly 30\%. Note that this quantity can be roughly interpreted as the integral under the blue band in Fig.~\ref{fig: Apogee vs. Bedell Slopes}, to the left of the black vertical dashed line at solar abundances. This is somewhat higher than the fraction of stars with negative [X/Fe] vs. $T_c$ slopes determined by \citet{2018ApJ...865...68B}, namely 19\% for abundances measured relative to hydrogen. 

\begin{figure}[tp!]
    \includegraphics[width=9cm]{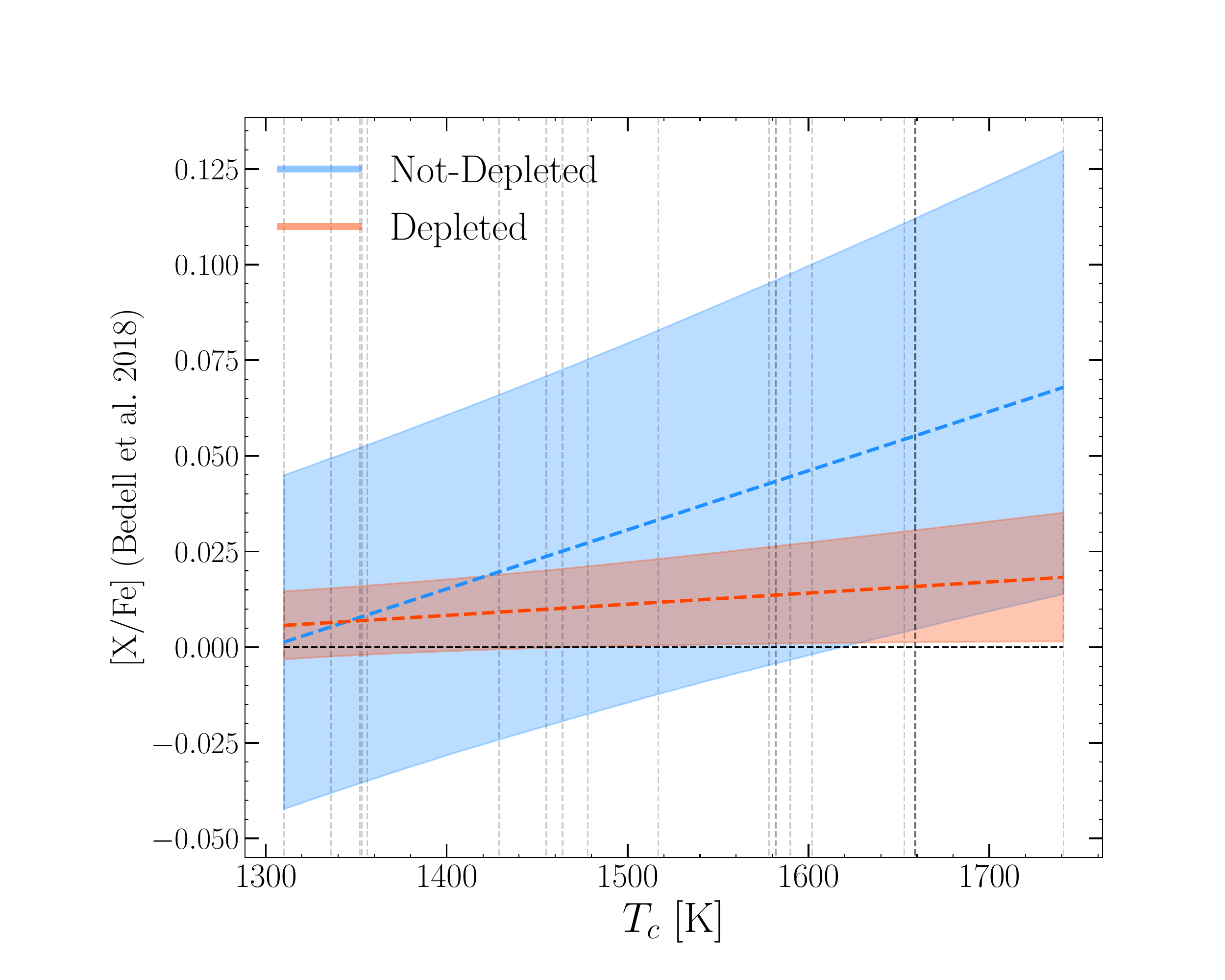}
    \centering
    \caption{Constraints on the depletion pattern for high precision abundance data from \citet{2018ApJ...865...68B} for a selection of 21 elements with $T_c > 1300~\rm{K}$. Dashed vertical lines are placed at the condensation temperature of each element, while the colored dashed lines depict the best fit mean depletion trend for D and ND (i.e. $\mu_m T_c + \mu_b$). }\label{fig:Bedell main result}
\end{figure}

Other studies report a more frequent occurrence of refractory depletions relative to the more volatile elements for various samples, thereby suggesting that the Sun's depletion pattern is not entirely uncommon. \citet{refId0}, for instance, finds that the fraction of stars with refractory to volatile element ratios as depleted as the Sun is in the range of $\sim 15\%$ for sub-solar metallicities to $\gtrsim 50\%$ for super-solar metallicities. Additionally, \cite{Gonzalez_2010} finds a significantly larger population of stars with such depletions ($\sim 50\%$) compared to \citet{2018ApJ...865...68B}, though their sample extends beyond solar analogs and specifically targets stars with planets regardless of stellar type. Evidently, larger sample sizes with high precision abundance measurements will be helpful to place more stringent constraints on the rarity of the solar depletion trend across various samples of stars. 

\subsubsection{Model Constraints on Data from Bedell et al. 2018}\label{sec: Bedell Results}

In this section, we present results from our model using high precision elemental abundance measurements of 79 solar analogs from \citet{2018ApJ...865...68B} and \citet{2018MNRAS.474.2580S} described in \S\ref{sec: Bedell Data}. The sample includes a similar population to the APOGEE stars described in \S\ref{sec: star selection}, with effective temperatures ($T_{\rm eff}$) generally within 100~K of solar, surface gravities ($\log{g}$) within 0.1~dex of solar, and metallicities (taken as [Fe/H]) within 0.1~dex of solar. While there are far fewer stars compared to our APOGEE sample, the number of available elements is significantly greater and the signal-to-noise for each abundance measurement generally higher. 

In order to allow for a more direct comparison to our constraints on APOGEE data in \S\ref{sec: APOGEE Results}, we use abundance measurements from \citet{2018ApJ...865...68B} relative to iron. Additionally, we include the 21 elements with $T_c > 1300~\rm{K}$ to more closely match the range of condensation temperatures used on APOGEE data. This also ensures the exclusion of moderately volatile elements (i.e. $900~\rm{K} \lesssim T_c \lesssim 1300~\rm{K}$), which in many cases \citet{2018ApJ...865...68B} finds to bias the refractory abundance vs. $T_c$ slope.

We choose not to employ outlier rejection in this section given the small sample size relative to wider surveys such as APOGEE. We believe this to be well motivated, since the line-by-line and strictly differential abundances from \citet{2018ApJ...865...68B} are less likely to have significant sources of systematic error within their sample \citep{2014ApJ...795...23B}. However, this does not ensure that systematics between abundance data from APOGEE and \citet{2018ApJ...865...68B} are necessarily reduced, since the use of varying stellar models, line lists, and generally different methodologies can complicate a straightforward comparison between the two studies.  

Our primary result for the 79 solar analogs and 21 elements are shown in Fig.~\ref{fig:Bedell main result}, where the abundance pattern is plotted in blue for ND stars and red for D stars. As before, vertical dashed lines are placed at the condensation temperature of each element. 
We find the mean slopes for ND and D stars to be significantly different, with
\begin{equation}\label{eq: bedell slope constraints}
    \mu_{m} = \begin{dcases*} 
         1.55 \pm ^{0.33}_{0.20}\times 10^{-4}~\rm{dex~K^{-1}}  \ \rm{for \ ND,} \\
         2.92 \pm ^{0.72}_{0.74}\times 10^{-5}~\rm{dex~K^{-1}} \ \rm{for \ D} \\
       \end{dcases*}
    \end{equation}\label{eq: mu_m constraints bedell}
at $68\%$ confidence. Our constraint on the fraction ($f$) of stars drawn from the depleted slope distribution for this sample is
\begin{equation}\label{eq: Bedell f constraint}
    f = 0.56 \pm 0.06,
\end{equation}
also at $68\%$ confidence. A systematic error bar is not provided on this constraint as was done in \S\ref{sec: APOGEE Results}, since the \citet{2018ApJ...865...68B} sample includes many more elements and constraints are found to be generally robust to small variations on which elements are included. 

The mean slopes are again in general agreement with measurements of individual stars  \citep[i.e.,][]{Mel_ndez_2009,refId0,2015A&A...579A..20M,2018ApJ...865...68B}, with a larger scatter compared to our constraints on APOGEE data in \S\ref{sec: APOGEE Results} which is due to a combination of both higher statistical uncertainty and a preference for larger intrinsic scatter.   
In addition, we provide contour plots of the model posterior for the main parameters of interest in Appendix~\ref{app:posterior on model parameters}.

Differences in the intrinsic scatter on the slopes ($\sigma_m$) and intercepts ($\sigma_b$) between our APOGEE constraints and the \citet{2018ApJ...865...68B} constraints are likely driven by the inclusion of a significantly larger number of elements in the \citet{2018ApJ...865...68B} sample. Furthermore, inaccuracies in the reported elemental abundance uncertainties can have the propagated effect of inflating or deflating the model fit for $\sigma_m$ and $\sigma_b$, since these parameters are meant to encode intrinsic scatter rather than measurement uncertainties. Small differences in intercepts and the absolute slopes are also to be expected, since APOGEE calibrates to a ``solar-like" scale whereas the \citet{2018ApJ...865...68B} measurements are conducted line-by-line, strictly differential to the Sun. Nonetheless, we find similar abundance vs. $T_c$ trends for both the D and ND model components between the two data sets.

A unique feature of our analysis compared to previous studies is our ability to place a constraint on the fraction ($f$) of stars drawn from the depleted slope distribution. For the APOGEE and \citet{2018ApJ...865...68B} sample, our constraint on this fraction is expressed in Eq.~\ref{eq: f constraint} and Eq.~\ref{eq: Bedell f constraint}, respectively. We find that when incorporating the systematic error bar provided on the APOGEE result in Eq.~\ref{eq: f constraint}, the \citet{2018ApJ...865...68B} constraint is consistent to within $\sim 1\sigma$.

As we find on APOGEE data, while the fraction of stars drawn from the depleted distribution is significant ($\sim 50\%$), the number of stars with depletion trends as extreme (i.e. as refractory depleted relative to the volatiles) as the Sun is comparably smaller ($\sim 20\%$). Perhaps unsurprisingly, this is consistent with \citet{2018ApJ...865...68B}, who finds that $19\%$ of stars have refractory to volatile ratios as refractory depleted as the Sun using abundances measured relative to hydrogen.

\subsection{Solar Depletion Trend and the Average Solar Analog}\label{sec: average solar analog}

Several studies have demonstrated that the Sun appears depleted in refractory elements relative to the majority of nearby solar analogs, with these results being extended to other planet host stars in some cases \citep{Mel_ndez_2009,refId0,2010A&A...521A..33R,Gonzalez_2010,2011ApJ...732...55S,2016MNRAS.456.2636L, 2018ApJ...865...68B,2020MNRAS.495.3961L}. In this section, we closely follow the methodology of \citet{2018ApJ...865...68B} and calculate the abundance trend of the ``average" solar analog compared to the Sun using our constraints on all model parameters derived from our APOGEE sample. In particular, the sample average is taken in linear space with the number density of atoms, so that the mean abundance ratio of element X to iron for a population of $\rm{N}$ stars is
\begin{equation}\label{eq: avg solar twin}
    \bigg\langle \rm{\left[\frac{X}{Fe}\right]} \bigg\rangle_{\rm Stars} = \log_{10}\left(\frac{1}{N}\sum_{n=0}^{N} 10^{\rm{[\frac{X}{Fe}]}_n}\right).
\end{equation}

\begin{figure}[tp!]
    \includegraphics[width=8.5cm]{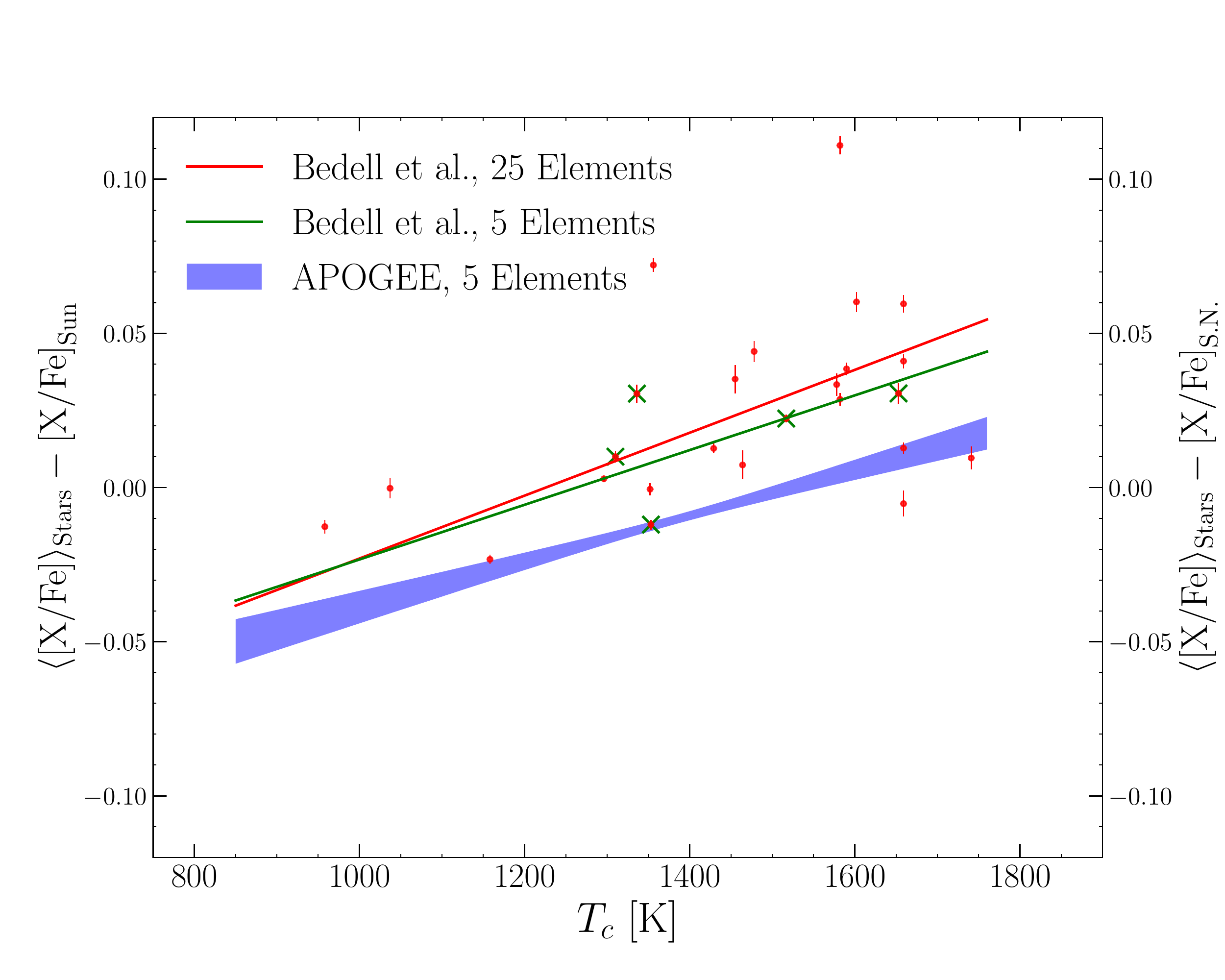}
    \centering
    \caption{The abundance pattern of the average solar analog compared to the Sun, using our constraints from APOGEE data (purple). Red and green points depict high precision abundance determinations of 79 solar analogs from \citet{2018ApJ...865...68B}. We fit linear least squares trends to all 25 elements plotted (red) and the five elements we study on APOGEE data (green). We reproduce the finding that the Sun is depleted in refractory elements relative to the more volatile elements. }\label{fig: avg solar twin}
\end{figure}

In order to derive the average abundance ratios from our results on APOGEE data, we simulate data with underlying parameters governed by our model constraints, computing Eq.~\ref{eq: avg solar twin} at the relevant condensation temperatures corresponding to each element. The resulting average abundance trend is illustrated in Fig.~\ref{fig: avg solar twin}. The width of the blue error band is derived from the 68\% CL on the average abundance as a function of $T_c$. For comparison, we also include linear fits to the average elemental abundance measurements from \citet{2018ApJ...865...68B} for a sample of 79 Sun-like stars in red. The green line depicts a linear fit to the same data, but limited to the five elements Si, Mg, Ni, Ca, Al (i.e. those used on APOGEE data). We include two y-axes in Fig.~\ref{fig: avg solar twin} to emphasize that the abundances from \citet{2018ApJ...865...68B} and APOGEE are measured on slightly different scales. In particular, abundance measurements from \citet{2018ApJ...865...68B} are measured relative to the Sun ([X/Fe]$_{\rm Sun}$), while abundances from APOGEE are measured relative to the solar neighborhood ([X/Fe]$_{\rm S.N.}$). Consequently, the offset in Fig.~\ref{fig: avg solar twin} between our APOGEE results and those from \citet{2018ApJ...865...68B} are likely driven by differences in the abundance scales.

 We find our constraints are consistent with the \citet{2018ApJ...865...68B} measurements up to an offset term.
 Furthermore, we recover the previously observed depletion trend, with the Sun appearing increasingly deficient in the high $T_c$ refractory elements relative to the lower $T_c$, more volatile elements.

We note that the result presented in this section is indeed consistent with our discussion in \S\ref{sec: APOGEE Results}, where we find that the Sun, itself, belongs to a majority distribution of stars with similar abundance vs. $T_c$ trends. The existence of a second population of stars (ND component) with refractory enhanced abundance trends relative to the more volatile elements, drives the elemental abundances of the average solar analog to higher values as a function of $T_c$. This leads to the upward sloping result in Fig.~\ref{fig: avg solar twin}. Consequently, our constraints are consistent with the claim that the Sun appears increasingly deficient in higher $T_c$ elements relative to the average solar analog, though our interpretation of this result differs. We discuss our interpretation of these results more thoroughly in \S\ref{sec: discussion}.

\section{Discussion}\label{sec: discussion}
\subsection{Summary}

We have developed a novel statistical approach to probe stellar elemental abundances as a function of element condensation temperature, $T_c$.
Using a two component mixture model, we fit the abundance measurements of roughly 1700 stars from APOGEE.  By combining constraints from many stars, we find evidence for two populations of stars: one population has an abundance vs. $T_c$ relation roughly in line with the Sun, while the other population has an enhanced abundance of refractory elements with high $T_c$.  We refer to these groups as the depleted and not-depleted populations, respectively.  The key advantage of our approach is that it can be applied to datasets for which the individual abundance measurements may have low signal-to-noise.  By combining constraints from multiple stars at the likelihood level, we can obtain a high significance constraint from a large sample of individually low signal-to-noise measurements. 

We find that the pattern of refractory element abundances seen in the Sun is not uncommon.  Indeed, the fraction of depleted stars (of which the Sun is a member) is found to be $f = 84\pm^{5\%}_{17\%}$ (including both statistical and systematic error) in the APOGEE sample.  The distribution of abundance vs. $T_c$ slopes inferred from our analysis is shown with the blue band in Fig.~\ref{fig: Apogee vs. Bedell Slopes}. 

These results can be compared to a recent analysis of similar abundance vs. $T_c$ trends by \citet{2018ApJ...865...68B}.  In contrast to our analysis, \citet{2018ApJ...865...68B} used a much smaller sample of stars (79 solar analogs), with more accurately measured abundances, and for a wider range of elements.  Our results appear to be in qualitative agreement with those of \citet{2018ApJ...865...68B}, as seen in Fig.~\ref{fig: Apogee vs. Bedell Slopes} and Fig.~\ref{fig: avg solar twin}.  When we apply our analysis methodology directly to the \citet{2018ApJ...865...68B} measurements, we also find fairly consistent results, as discussed in \S\ref{sec: Bedell Results}.

However, our interpretation of these results differs somewhat from that of \citet{2018ApJ...865...68B}.  \citet{2018ApJ...865...68B} note that the Sun shows a greater depletion of refractory elements compared to roughly 80\% of stars.  We find a similar result in our analysis; this fraction is roughly equivalent to the integral under the blue band in Fig.~\ref{fig: Apogee vs. Bedell Slopes} below the black dashed line. 
However, our interpretation is not that the Sun is unusually depleted in refractory elements, but rather that it sits close to the middle of a distribution of similar, refractory-depleted stars that make up a large portion of all stars.

\subsection{Implications for abundance of small planets}\label{sec: implications small planets}
\citet{Mel_ndez_2009} were the first to claim that the Sun has a deficiency in refractory elements relative to volatiles compared to other Sun-like stars. This result has since fostered an ongoing debate, with some works verifying the Sun's unusual abundance pattern and others refuting it \citep{refId0,2014A&A...564L..15A,2015A&A...579A..52N,2018ApJ...865...68B}.  While we find evidence that the Sun shows a greater depletion than the average solar-like star, as noted above, our analysis suggests that the Sun may actually be a part of a dominant population of stars that show similar trends.

If we adopt the rocky planet hypothesis proposed by \citet{Mel_ndez_2009}, our results can be interpreted as separating out stars with rocky planets from those without, with roughly 80\% belonging to the former group. 
Some arguments against the rocky planet hypothesis stem from the fact that few stars seem to have refractory depletion trends as extreme as the Sun, while far more stars likely host small rocky planets  \citep{2011ApJ...737L..32S,2015ApJ...809....8B,2020MNRAS.493.5079B}. Our results suggest that the Sun actually falls into a dominant population of refractory-depleted stars.

One can further speculate that the placement of a star in the distribution of abundance vs. $T_c$ slopes is determined not solely by its rocky planet hosting status, but may also be related to giant planet formation. Recently, \citet{2020MNRAS.493.5079B} argue this to be the case, since they demonstrate that a forming giant planet can create a pressure trap in the protoplanetary disk, thereby inhibiting refractory material from accreting onto the host star. They make the argument that rocky Earth-like planets are not the likely culprits of the solar depletion pattern, since rocky planets are far more common \citep{2003ApJ...598.1350L,2011ApJ...742...38Y,2013ApJ...767L...8K,2019AJ....158..109H} (as discussed above, our interpretation of the depleted population does not support this argument, given the large population of stars with near solar abundances). Their results also suggest that there is not enough refractory material locked up in the Earth to account for the observed solar depletion trend, making the giant planet hypothesis more appealing. 

Adopting the planet hypothesis for refractory depletions, our approach can in principle address both the rocky planet explanation proposed by \citet{Mel_ndez_2009} and the giant planet hypothesis previously discussed \citep{2020MNRAS.493.5079B}. Because the detection rate of giant planets is low ($\sim 15\%$; \citealt{2019ApJ...874...81F,2016ApJ...819...28W,2020MNRAS.492..377W}), some stars may reside towards the more depleted tail of the abundance vs. $T_c$ slope distribution. In particular, stars like the Sun, with multiple terrestrial and gas giant planets, may have excess depletions compared to the typical depleted star if both types of planets are associated with depletion. We find that the Sun is in fact over $1\sigma$ depleted relative to the mean of the depleted population. Significant additional work is needed to explore and validate these correlations.

Regardless of the hypothesis one adopts, we suggest that future works studying subtle elemental depletion patterns in various samples of stars more rigorously test for the presence of bimodalities in the distribution of abundance vs. $T_c$ slopes. Doing so provides a more complete picture of where the Sun falls in its elemental composition relative to other stars, paving the way to a more thorough understanding of the planet-star connection. 

\subsection{Caveats}\label{sec: caveats}
Galactic chemical evolution effects (GCE) are not corrected for in this analysis, and could lead to elemental abundance trends with $T_c$ \citep{2014A&A...564L..15A,2016A&A...593A.125S}. \citet{2018ApJ...865...68B} addresses this by fitting age trends to the abundances of solar analogs in their sample as a function of inferred stellar age, thereby indirectly probing GCE. They find that when subtracting off the inferred GCE fit from all abundances, the Sun still appears increasingly depleted in higher $T_c$ elements relative to the average solar analog, suggesting that GCE alone is not driving the refractory depletion trends. Larger samples that include more elements with similar chemical origins will help constrain the extent to which GCE could be responsible for elemental abundance trends with $T_c$.

With respect to our analysis of APOGEE data, binary star systems can adversely effect chemical abundance measurements either by having changed the evolution of the star targeted by APOGEE or, if spectral features from both stars can be identified from the spectrum, by changing the spectral features being measured. The APOGEE observation strategy was designed to be sensitive to close binaries \citep[see discussion in][]{2017AJ....154...94M}.
 By modeling the multi-epoch radial velocities of APOGEE DR16 stars,  \citet{2020ApJ...895....2P} generated a catalog of roughly $20000$ close binary companions enabling us to crosscheck our sample for possible binary contamination.\footnote{The results of this analysis are available as a DR16 Value Added Catalog here \url{https://www.sdss.org/dr16/data_access/value-added-catalogs/?vac_id=orbital-parameter-samplings-of-apogee-2-stars-from-the-joker}} We find that $\sim 2\%$ of stars in our final sample are potential binaries from the \citeauthor{2020ApJ...895....2P} analysis. Since both D and ND components prefer fractions greater than 10\%, binary star systems are unlikely to be the driving force behind the observed shifting bimodalities. Indeed, after removing the potential binaries from our sample constraints remain virtually unchanged. 
An alternate strategy presented in \citet{2018MNRAS.476..528E} models the spectra, themselves, rather than using RV-variations to identify likely binaries. This approach is sensitive to a slightly different population of binaries; the binary-fractions, however, remain small compared to the D and ND signals in our work.

Due to calibrations differences between the APOGEE elemental abundances and other works, it is not necessarily valid to compare the absolute abundance measurements from APOGEE to those derived from a strictly differential analyses measured relative to the Sun (i.e., as in \citealt{2018ApJ...865...68B}). In \S\ref{sec: APOGEE Results}, however, we find that the APOGEE solar reference spectrum measured in reflected sunlight from the asteroid Vesta \citep{2020arXiv200705537J} is roughly consistent with an abundance vs. $T_c$ slope and intercept of zero on the APOGEE scale. Moreover, the solar pattern is consistent with the depleted population.  This is further validated by our results in \S\ref{sec: distribution of abundance vs. Tc slopes}, where we find that our constraints from APOGEE yield similar elemental abundance patterns to those from \citet{2018ApJ...865...68B}, albeit with some offsets (with respect to the not-depleted abundance vs. $T_c$ slopes and intercepts, in particular). Consequently, while calibration offsets could have the effect of shifting absolute elemental abundance trends, we do not believe this to be the dominant factor in the trends constrained by the APOGEE abundance data. Furthermore, because APOGEE calibrations are simply applied as a zero-point offset per element to all stars, the presence or absence of distinct populations in the distribution of elemental abundances is unaffected, as well as the relative slopes between the depleted and not-depleted model components.

In order to achieve a high level of precision, our analysis on APOGEE data is limited to the five elements Si, Mg, Ni, Ca, Al, spanning a range of condensation temperatures from $\sim 1300$--$1600~\rm{K}$. Consequently, our constraints should be interpreted as applying only to this range of $T_c$, since some stellar abundance trends are better described by piecewise linear functions when incorporating elements spanning a wider range of condensation temperature \citep{Mel_ndez_2009,2016A&A...588A..81S,2018ApJ...865...68B}. Increasing the number of high precision, high $T_c$ elements will be useful for future studies of elemental depletion patterns.

Our constraints from APOGEE presented in \S\ref{sec: APOGEE Results} are potentially sensitive to systematic uncertainties stemming from the ASPCAP pipeline. While we have chosen elements with small calibration zero-point shifts to the mean abundances in the solar-neighborhood  ($\lesssim 0.04~\rm{dex}$), inaccuracies in abundance determinations could bias model constraints. In particular, we find evidence of two populations of stars; one with roughly solar abundances and another which is increasingly enhanced in refractory elements as a function of $T_c$, indicating a departure from the solar neighborhood. Because APOGEE calibrates to the solar neighborhood, stars which depart from this trend (i.e. the not-depleted population) could be over corrected for in the ASPCAP pipeline. We believe this to be the underlying reason why the not-depleted constraints on APOGEE prefer a relatively low intercept and high slope in Fig.~\ref{fig:main result} compared to our results from \citet{2018ApJ...865...68B} in Fig.~\ref{fig:Bedell main result}. Regardless, the relative abundance trends between depleted and not-depleted populations in both results are similar.

Relevant to the results of this work, if the true uncertainties on the APOGEE abundances are higher than reported, the intrinsic scatter model parameters (i.e. $\sigma_m, \   \sigma_b$, etc.) will account for both intrinsic dispersion and noise fluctuations. Our model constraints on all intrinsic scatter parameters therefore rely on the accuracy of the APOGEE provided uncertainties, and those provided by \citealt{2018ApJ...865...68B} (and \citealt{2018MNRAS.474.2580S} for [Fe/H]) in \S\ref{sec: Bedell Results}. We note, however, that inaccuracies in abundance uncertainties are not found to bias the main model parameters of interest on simulated data ($f, \ \mu_m, \ \mu_b$), since these parameters represent population means.
For further details on how the abundances and their associated uncertainties used in this analysis are derived, we direct readers to \citet{2020arXiv200705537J}, \citet{2018MNRAS.474.2580S}, and \citet{2018ApJ...865...68B}.

Our modeling scheme described in \S\ref{sec: modeling} assumes that the distribution of abundance vs. $T_c$ slopes and intercepts are Gaussian. For this reason, we impose outlier clipping on the abundance data to ensure that departures from gaussianity in the distribution tails are not driving the model fit. We adopt a conservative outlier rejection of $2\sigma$ on the distribution of abundances (see \S\ref{sec: outlier rejection} for details), and find that our main results are robust to variations on this threshold (Appendix~\ref{sec: variations on outlier rejection}). While the goodness of fit and tests for robustness suggest that our finding of two populations of stars is legitimate, larger sample sizes with more elements will help to further determine the validity of this claim.

Lastly, while it is true that the abundances in ASPCAP are determined independently, families of elements are not completely independent in that the underlying spectral parameters include abundance terms for [M/H] and [$\alpha$/M] by which individual abundance meausrements are constrained \citep[][]{2016AJ....151..144G}. However, the elements used in this analysis include two elements from the metal-family (Ni, Al) and three elements from the $\alpha$-family (Si, Mg, Ca). However, these classifications are not correlated with $T_c$ and are unlikely to drive the $T_c$ slopes. Moreover, the calibration to the solar-neighborhood is performed element-by-element and should take into account systematics in the individual element values.

\subsection{Future directions}

Our model enables the use of large, automated spectroscopic surveys with many stars to place constraints on the distribution of elemental abundances and subtle trends with condensation temperature.  Similar studies could also be conducted using data from the GALAH Survey or LAMOST \citep{2018MNRAS.478.4513B,2019ApJS..240....6Y}. 

Our sample is limited to a population of solar analog stars with similar stellar parameters to the Sun. Few analyses have explored beyond solar-like samples in order to achieve high precision abundance measurements and reduce stellar model-driven biases \citep{2018ApJ...865...68B}. \citet{Gonzalez_2010}, however, extended their sample to a wider range of effective temperatures and metallicities (taken as [Fe/H]), and found that super-solar metallicity stars with planets have the most refractory depleted abundance trends. \citet{refId0} also finds intriguing behavior at super-solar metallcities, with evidence of a bimodality in the distribution of [X/Fe] vs. $T_c$ slopes at 2.7$\sigma$. These findings are particularly interesting, given that the frequency of planets is widely thought to increase with stellar metallicity \citep{2005ApJ...622.1102F,2015AJ....149...14W}. Because our analysis relies on large samples of stars, it is well suited to using splits of the data to explore trends in depletion patterns with varying stellar properties.

Previous studies (i.e. \citealt{Gonzalez_2010}) have split their sample into stars with detected planets and stars without detected planets, evaluating the abundance vs. $T_c$ trend for each group. \citet{Gonzalez_2010} found that stars with planets tend to have more refractory depleted abundance trends than stars without detected planets, though other works have found no such correlation \citep{2013ASPC..472...97G,2015ApJ...815....5S}. Our analysis can be applied to stars with planets and without planets separately to directly test the connection of abundance trends with planet hosting status. Applying such analyses to large datasets from APOGEE, GALAH, and LAMOST will help contribute to our understanding of the nature of elemental depletion patterns with condensation temperature, and possible links to planet formation. 

\section*{Acknowledgments}
We are grateful to Megan Bedell and Cullen Blake for helpful comments on an earlier draft of the paper. We also thank Gary Bernstein, Neal Dalal, Mike Jarvis, Marco Raveri, Matthew Belyakov, Sarah Kane, Eli Wiston, and Maureen Iplenski for helpful discussions. 
We are grateful to Sten Hasselquist, Christian Hayes, and Matthew Shetrone for their invaluable insights regarding APOGEE data and APOGEE data analysis. 
JN has been supported in part by the Center for Undergraduate Research at the University of Pennsylvania. Further support for this work was provided by NASA through Hubble Fellowship grant \#51386.01 awarded to RLB by the Space Telescope Science Institute, which is operated by the Association of  Universities for Research in Astronomy, Inc., for NASA, under contract NAS 5-26555.

\facilities{Du Pont (APOGEE), Sloan (APOGEE), {\it Gaia}, NASA Exoplanet Archive}

\appendix
\restartappendixnumbering
\section{Star Selection and Abundance Distributions}\label{app: star selection and abundance distributions}
In this appendix we provide additional visualizations of the star selection described in \S\ref{sec: star selection} and the resulting distribution of elemental abundances as described in \S\ref{sec: element selection}.
\begin{figure*}[htp!]
    \includegraphics[width=16.5cm]{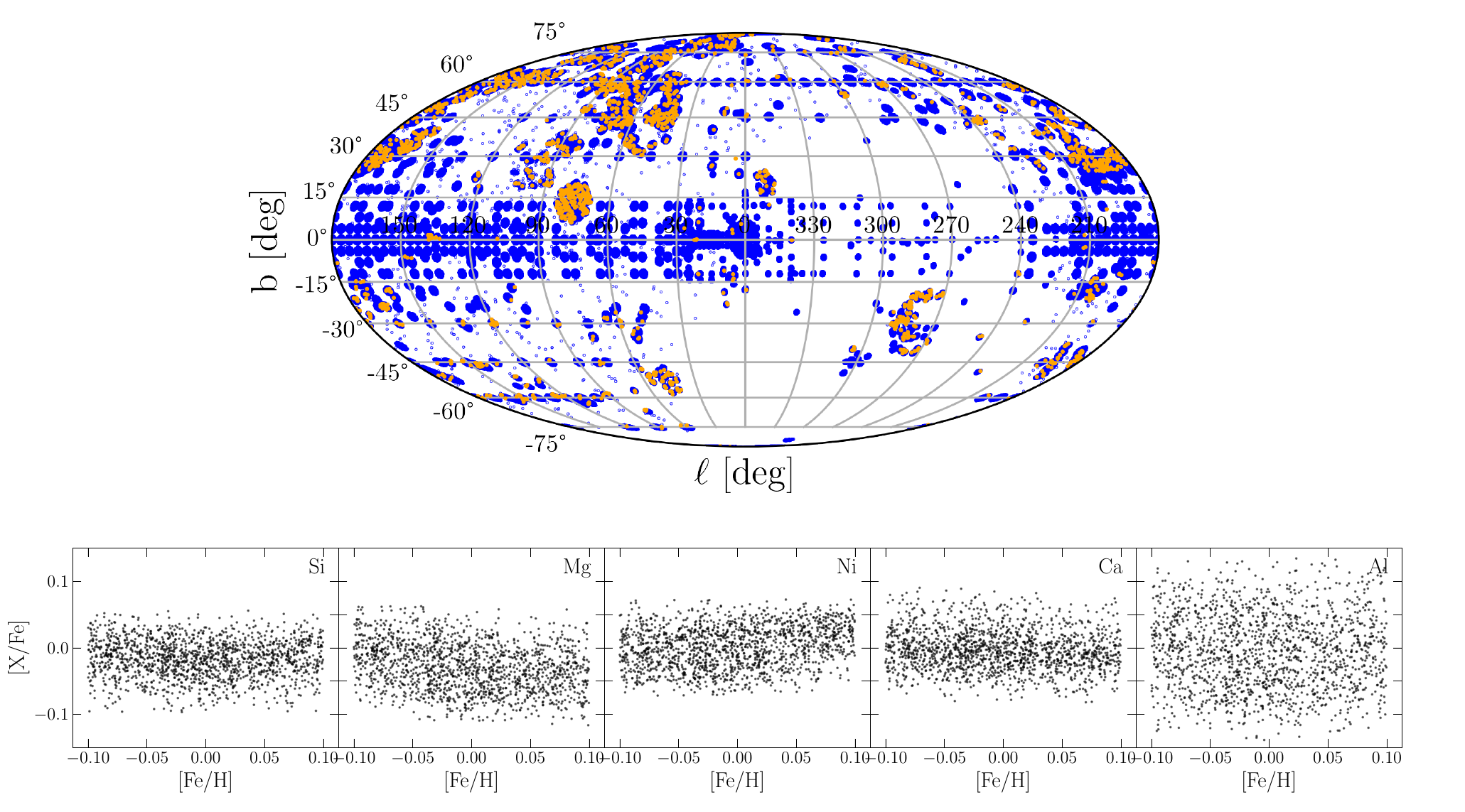}
    \centering
    \caption{Top: Mollweide projection of the APOGEE sample in Galactocentric coordinates. Orange points indicate the population of stars used in our analysis, described in \S\ref{sec: star selection}. Bottom: APOGEE abundance ratios relative to iron plotted against [Fe/H] for the population of Sun-like stars used in this work (described in \S\ref{sec: star selection}). These distributions show no evidence for $\alpha$-bimodality known to exist in the Galaxy for this sample of stars.}\label{fig:Mollweide and Abundace Distributions}
\end{figure*}

In Fig.~\ref{fig:Mollweide and Abundace Distributions} (top panel), we provide a Mollweide projection of our star sample in Galactocentric coordinates. Stars surviving our baseline cuts are plotted as orange points, while the rest of the APOGEE-2 sample are shown as blue points. 
Fig.~\ref{fig:Mollweide and Abundace Distributions} demonstrates that our sample is not drawn from a single sub-structure or sub-program in APOGEE, which reduces, if not eliminates, concern over sample bias. 
The majority of our stars are at high Galactic latitude ($|b|$ \textgreater 10$^{\circ}$) and this occurs because the APOGEE color-magnitude targeting algorithms are more likely to select solar-type dwarfs out of the Galactic plane where de-reddened color limits are relaxed to target bluer stars \citep{2013AJ....146...81Z,2017AJ....154..198Z}. 
Moreover, scientific programs within the APOGEE survey that specifically target dwarfs use fields out of the Galactic plane, like the {\it Kepler}-field and the {\it TESS} Continuous Viewing Zones (R.~Beaton et al., F.~Santana et al.~in prep.).

The Galaxy contains two clear chemical populations that are separated by their behavior in the [$\alpha$/Fe] and [Fe/H] diagram into an $\alpha$-high and $\alpha$-low population. 
The relative numbers of stars in either population varies as a function of location in the Galaxy, which is shown vividly by \citet[][their Fig.~4]{2015ApJ...808..132H}. 
The two populations, however, merge in the solar neighborhood to more-or-less Solar-like values.
\citet{2019ApJ...874..102W} specifically investigated chemical evolution trends in the two sequences and found, more or less, that the chemical evolution within the two $\alpha$-sequences appears to occur in a similar fashion (with exceptions for elements with more complicated yields); stated differently, the overall chemical trends within the populations are similar, but the $\alpha$-abundance is offset.
In \citet[][their Fig. 7]{2019ApJ...874..102W}, the nucleosynthetic pathways for the elements are summarized and from this the elements used in our analysis (Si, Mg, Ni, Ca, and Al) all have some component from core-collapse supernova that drive $\alpha$-element abundances at early times. 
Given our model looks for the presence of two populations and most of our elements have some core-collapse SN origins, it is not unreasonable to ensure that our sample shows no $\alpha$-bimodality that in turn could be driving the depletion signal.

\begin{figure*}[htp!]
    \includegraphics[width=16.5cm]{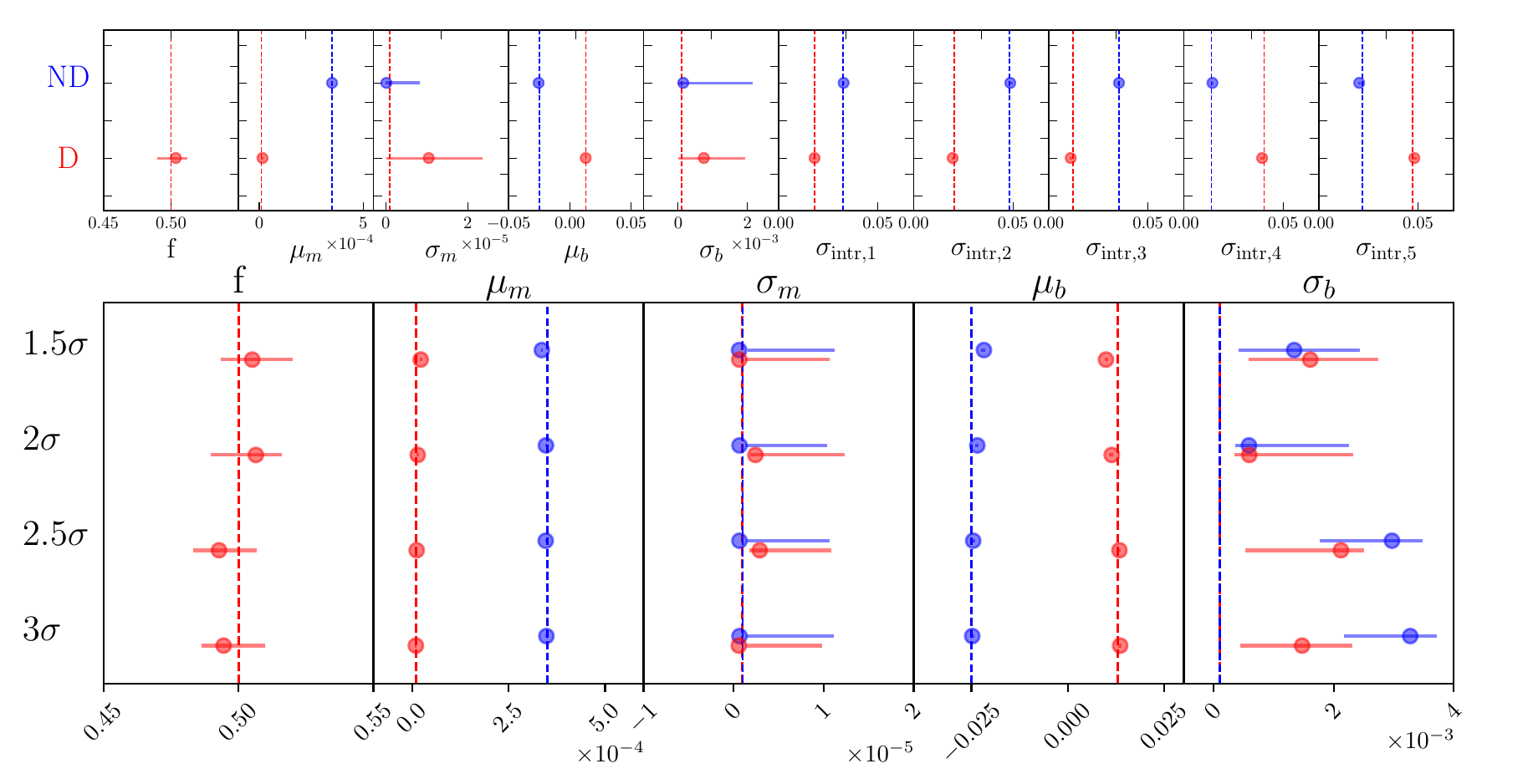}
    \centering
    \caption{Top: Results on simulated data for five elements with the same condensation temperatures used on actual data in \S\ref{sec: APOGEE Results}. ND (blue) and D (red) correspond to not-depleted and depleted stars, respectively. The blue and red vertical dashed lines represent the input parameters for the respective groups. Bottom: Results on simulated data with a Student t-distribution truth for various outlier rejection choices. Note that our model assumes normality, motivating the use of outlier rejection on actual data. Input parameters are given by the vertical dashed lines for depleted (red) and not-depleted (blue) stars. We omit constraints on the intrinsic scatter parameters ($\sigma_{{\rm intr},j}$) due to the high degree of degeneracy with the degree of freedom parameter $\nu$.  }\label{fig: simulation result}
\end{figure*}

In Fig.~\ref{fig:Mollweide and Abundace Distributions} (bottom panel), we plot abundances relative to iron against [Fe/H] for the five elements from APOGEE data used in this analysis. Points are only plotted for stars passing our analysis cuts. We do not find clear evidence of multiple modes for any of the elements in Fig.~\ref{fig:Mollweide and Abundace Distributions}, indicating that our model fit should not be driven by two populations of stars with vastly different chemical origins. There are slight trends for Mg and Ni over the full range of [Fe/H], but these trends are not likely to drive the conclusions found in the main text.

\section{Analysis of Simulated data}\label{app: analysis of sim data}

Constraints on simulated chemical abundance data for five elements and 2000 stars with uncertainties sampled directly from APOGEE are provided in the top panel of Fig.~\ref{fig: simulation result}. A more complete description of simulation procedures is provided in \S\ref{sec: Analysis of Simulated Data}. Red and blue lines correspond to the true parameters values for depleted (D) and not-depleted (ND) stars, respectively. We note that the abundance vs. $T_c$ intercept in Fig.~\ref{fig: simulation result} is defined at $T_c = 1300~\rm{K}$ rather than zero in order to reduce degeneracy with $\mu_m$. We find that our analysis pipeline is successful in recovering the true parameters on simulated data for various choices of input parameters. 

\subsection{Departures from Gaussian Assumption}\label{sec: Departures from Gaussian Assumption} 
In order to evaluate how reasonable departures from our Gaussian assumption impacts model constraints, we generate abundance data from the Student's t-distribution. The Student's t-distribution is symmetric and bell-shaped like the normal distribution, but can have significantly heavier tails depending on the number of degrees of freedom ($\nu$) that one assumes. This means that the distribution is more likely to produce values farther from the mean, providing us with a useful test of our outlier rejection scheme. 

We simulate data by generating mock elemental abundances from the Student's t-distribution with $\nu = 2$ degrees of freedom, giving significant weight to the tails of the abundance distributions. We suppose that all abundance vs. $T_c$ slopes and intercepts are normally distributed as in \S\ref{sec: Analysis of Simulated Data}, with only the resulting distribution of abundances departing from Gaussian.

Results for this test are presented in the bottom panel of Fig.~\ref{fig: simulation result}, with rejection thresholds ranging from $1.5\sigma$ to $3\sigma$. We omit constraints on the intrinsic scatter parameters $\sigma_{{\rm intr},j}$ due to the high degree of degeneracy with the Student t parameter $\nu$. For this reason, parameter constraints on $\sigma_{{\rm intr},j}$ are not emphasized in this analysis, since a poor fit to the intrinsic scatter on both simulated and actual data will become evident in the goodness-of-fit test described in Appendix~\ref{app: goodness of fit}.

We find the most significant bias in $\sigma_b$ for less conservative outlier rejections (i.e. for $\geq 2.5\sigma$), but note that constraints are generally robust across all rejection choices. We therefore adopt a rejection threshold of $2\sigma$ on APOGEE data for our fiducial result, since this choice provides a conservative rejection with minimal bias in the key parameters of interest ($f, \ \mu_m, \ \mu_b$ for depleted (red) and not-depleted (blue) stars). We explore variations on the fiducial outlier rejection threshold on actual data in Appendix~\ref{sec: Robustness of Apogee constraints}.

\section{Testing the Goodness of Fit}\label{app: goodness of fit}
\begin{figure*}[htp!]
    \includegraphics[width=16.5cm]{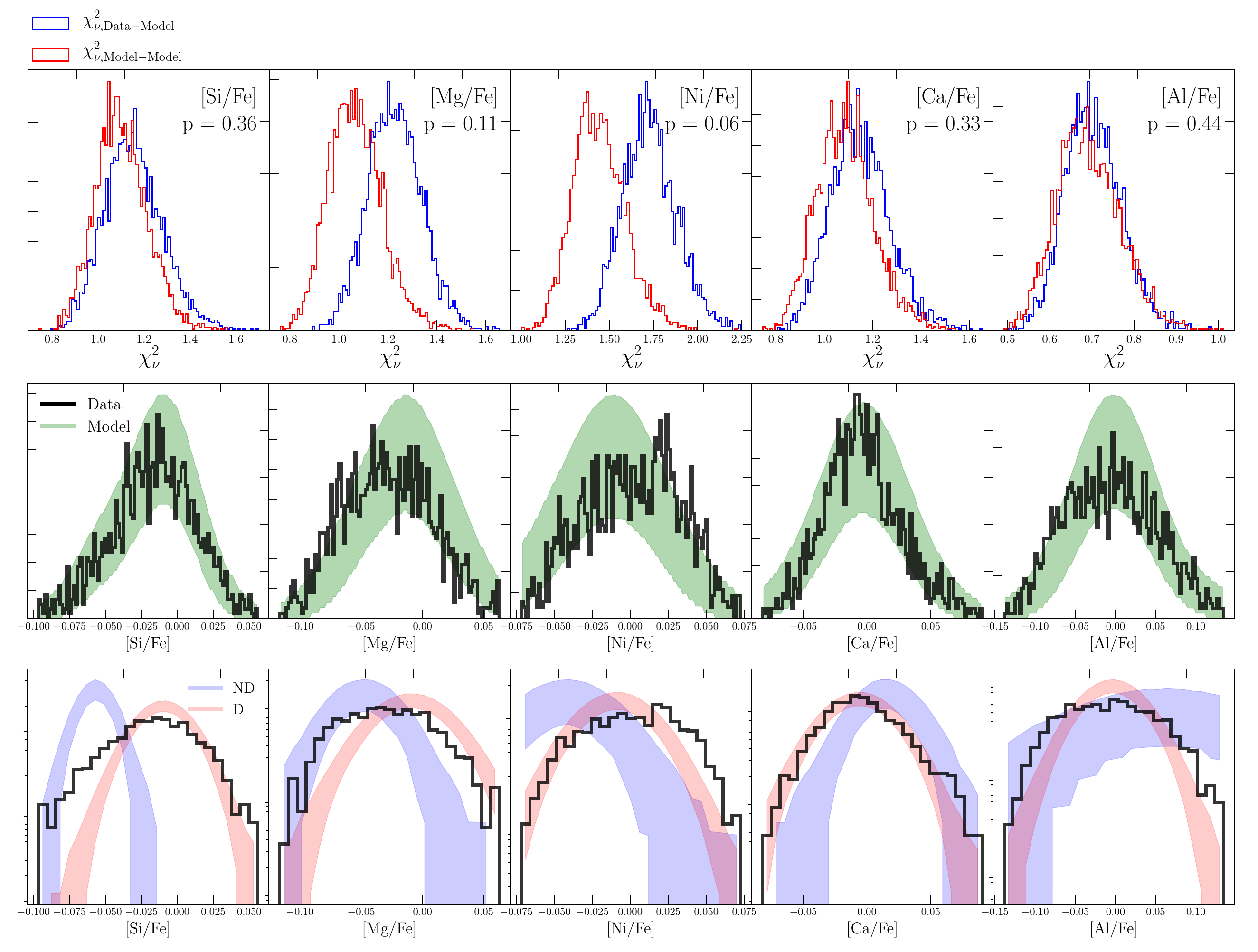}
    \centering
    \caption{Top Panel: Posterior predictive distributions for the reduced $\chi^2$ between the data and model (blue) and two simulated realizations of the model (red). The p-value is defined as the fraction of times the $\chi^2$ between the data and the model is less than the same metric between the two simulated realizations of the model. Middle Panel: Visualization of the model fit to data, where the x-axis is the elemental abundance [dex] and the y-axis is a histogram count. The black histogram shows the distribution of [X/Fe] for APOGEE data, while the green histograms depicts the 68\% CL from simulated data with underlying parameters governed by the MCMC chains. Bottom Panel: Same as the middle panel, but with the two components (ND and D) plotted separately. Note that the y-axis is plotted on a logarithmic scale.}\label{fig:PPD APOGEE 5 elements}
\end{figure*}

In order to assess the quality of the model fit, we make use of a posterior predictive check, described as follows. Let $\boldsymbol{\theta_i}$ be the vector of model parameters corresponding to the $i^{th}$ step of the MCMC parameter chain after appropriate burn-in. For each step in the parameter chain, we simulate data according to the modeling outlined in \S\ref{sec: modeling}, with $\boldsymbol{\theta_i}$ as the underlying model parameters. We then histogram the simulated data with the same bins as the real, observed data from APOGEE. The result is that we are able to compare the simulated data to the observed data for each step in the chain, while also making the same comparison between two simulated realizations of data.

In order to quantify differences between the simulated data realizations and the actual data, we make use of the reduced-$\chi^2$ (denoted $\chi^2_{\nu}$) test statistic. Let $n^{\rm obs}_{i}$ be the counts of the observed data in bin $i$ for element X, while $n^{\rm sim}_i$ represents the same quantity but for simulated data. We then define the reduced-$\chi^2$ between the data and the model as

\begin{equation}
    \chi^2_{\nu, \rm{Data - Model}} \equiv \frac{1}{N-1}\sum_i \frac{\left(n^{\rm obs}_{i} - n^{\rm sim}_{i}\right)^2}{n^{\rm sim}_{i}},
\end{equation}
where $N$ is the number of stars and $n_{\rm sim}$ is assumed to be Poisson-distributed. We also compute the same statistic between two simulated realizations of the model with underlying parameters $\boldsymbol{\theta_i}$. In this case, the reduced-$\chi^2$ is
\begin{equation}
     \chi^2_{\nu, \rm{Model - Model}} \equiv \frac{1}{N-1}\sum_i \frac{\left(n^{\rm sim*}_{i} - n^{\rm sim}_{i}\right)^2}{n^{\rm sim}_{i}},
\end{equation}
where $n^{\rm sim*}_{i}$ is the histogram counts for an additional realization of the model with underlying parameters $\boldsymbol{\theta_i}$. 

The general logic is that if the model provides a decent fit to the data, the distributions of $\chi^2_{\nu, \rm{Data - Model}}$ and $\chi^2_{\nu, \rm{Model - Model}}$ should be similar, with  $\chi^2_{\nu, \rm{Data - Model}}$ occasionally indicating a better fit than $\chi^2_{\nu, \rm{Model - Model}}$.

\subsection{Goodness of Fit: Data from APOGEE}\label{app: APOGEE PPD five elements}

We provide the results of the posterior predictive test in Fig.~\ref{fig:PPD APOGEE 5 elements} (top panel), where we plot $\chi^2_{\nu, \rm{Data - Model}}$ in blue and $\chi^2_{\nu, \rm{Model - Model}}$ in red for each element. The p-value in this context does not match the frequentist definition, but is defined as the fraction of times $\chi^2_{\nu, \rm{Data - Model}}$ is less than $\chi^2_{\nu, \rm{Model - Model}}$. Thus, ``perfect" p-values are expected to fall around 0.5. With the exception of Ni, all p-values are greater than 0.10 indicating a generally strong fit. The overall p-value, determined by summing over all elements is 0.06. We find that when removing Ni abundances and running the analysis again, parameter constraints remain consistent with the case of all five elements included. Furthermore, the overall p-value with Ni removed is p = 0.32, indicating an excellent fit to the data. While the overall p-value is significantly better after removing Ni from the analysis, the consistency of model constraints with Ni removed compared to the case of all five elements included indicates that Ni alone is not driving the model fit in \S\ref{sec: APOGEE Results}. 

The middle panel of Fig.~\ref{fig:PPD APOGEE 5 elements} provides a visualization of the model fit to APOGEE data for the five elements Si, Mg, Ni, Ca, Al. The black curve is a histogram of the abundance data for each element over our solar analog sample, while the green band is derived from our model constraints. We note that generating the green band in Fig.~\ref{fig:PPD APOGEE 5 elements} makes use of all model parameters providing a useful visualization of the fit. Meanwhile, the strength of the fit is quantified in the top panel. 

The bottom panel of Fig.~\ref{fig:PPD APOGEE 5 elements} provides a further visualization of our model constraints presented in \S\ref{sec: APOGEE Results}. In particular, we plot the depleted and not-depleted model components separately in Fig.~\ref{fig:PPD APOGEE 5 elements}, over the distribution of elemental abundances (in black) from our solar analog sample. The mean of the depleted population can be seen to remain roughly consistent with [X/Fe] = 0 across the board, while the ND component shifts from left to right for higher $T_c$ elements. We note that the amplitudes of the D and ND components in this figure are not scaled by the parameter $f$ (or $1-f$) in order to place the two distributions on the same scale. 

\subsection{Goodness of Fit: Data from Bedell et al. 2018}\label{app: Bedell PPD 25 elements}

In this appendix we evaluate the goodness of fit of the model constraints on abundance data from \citet{2018ApJ...865...68B} of 79 stars and 21 elements. Given the large number of elements in this section, we report an overall p-value in place of individual element p-values. We find that the model provides an extremely good fit to the data, with an overall p-value of 0.27. 

\section{Robustness of \normalfont{APOGEE} Constraints}\label{sec: Robustness of Apogee constraints}

In this appendix we test the robustness of our fit to APOGEE data for the five elements Si, Mg, Ni, Ca, Al. 
\subsection{Variations on Outlier Rejection}\label{sec: variations on outlier rejection}
\begin{figure*}
    \includegraphics[width=16.8cm]{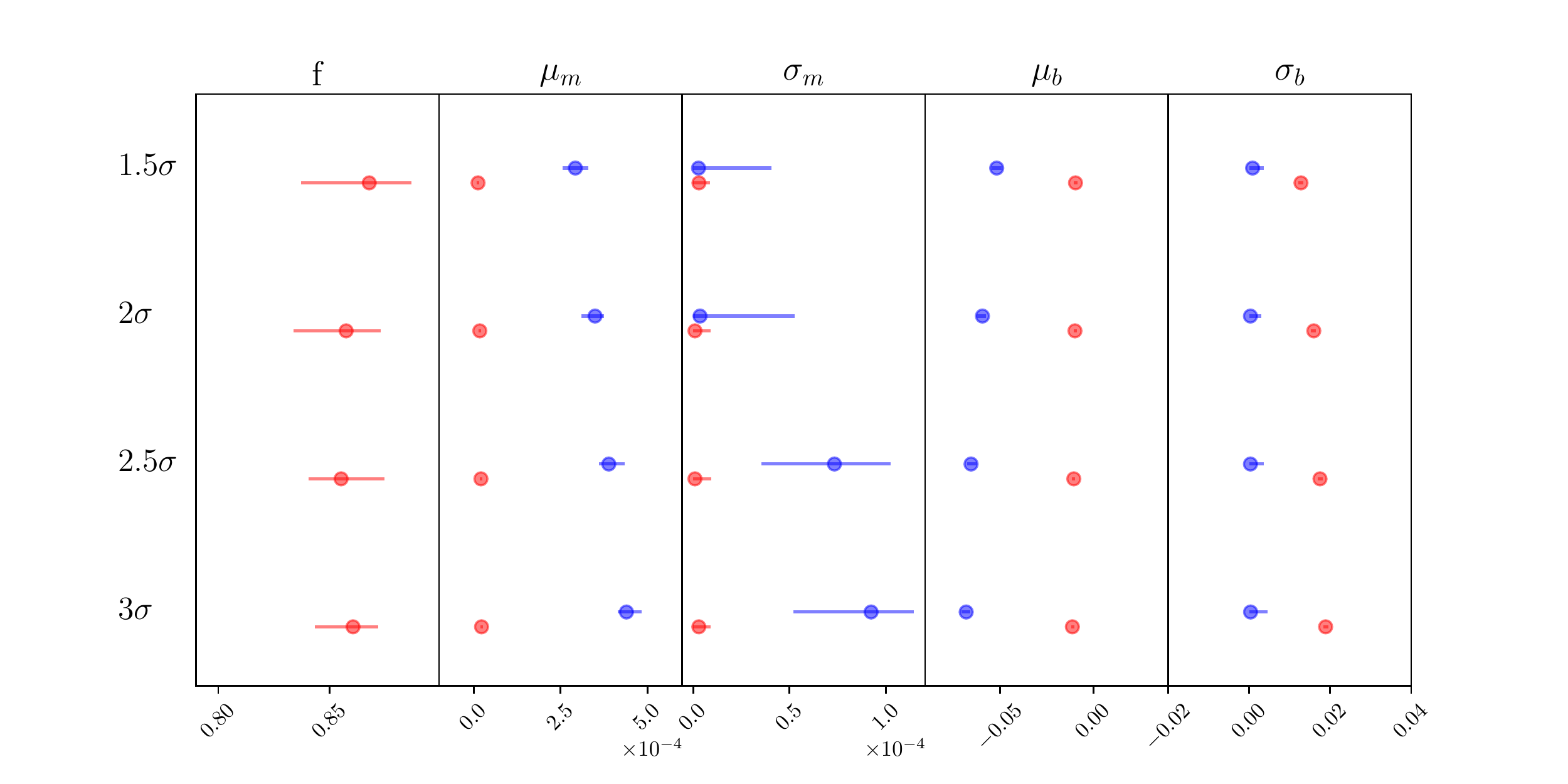}
    \centering
    \caption{Results on APOGEE data for various outlier rejection thresholds. Red and blue points correspond to D and ND model components, respectively. We find that for variations around the $2\sigma$ rejection threshold adopted in \S\ref{sec: APOGEE Results}, parameter constraints are robust. }\label{fig: outlier reject APOGEE data}
\end{figure*}

In \S\ref{sec: outlier rejection} we discuss the outlier rejection scheme used on APOGEE data, and present variations on the rejection threshold in this section. Our fiducial result presented in \S\ref{sec: APOGEE Results} adopts a $2\sigma$ rejection threshold, such that any star with $\abs{\rm{[X/Fe]}-\langle\rm{[X/Fe]}\rangle} > 2\sigma \left(\rm{[X/Fe]} \right)$ is removed from the analysis (see \S\ref{sec: outlier rejection}). In Fig.~\ref{fig: outlier reject APOGEE data}, we provide model constraints on the main parameters of interest for various rejection thresholds ranging from $1.5\sigma$ to $3\sigma$. Indeed, our constraints are robust to variations on the rejection threshold, indicating a robust fit to the data.

\subsection{Element Exclusion Test}\label{sec: element exclusion test}

\begin{figure*}[ht!]
    \includegraphics[width=16.8cm]{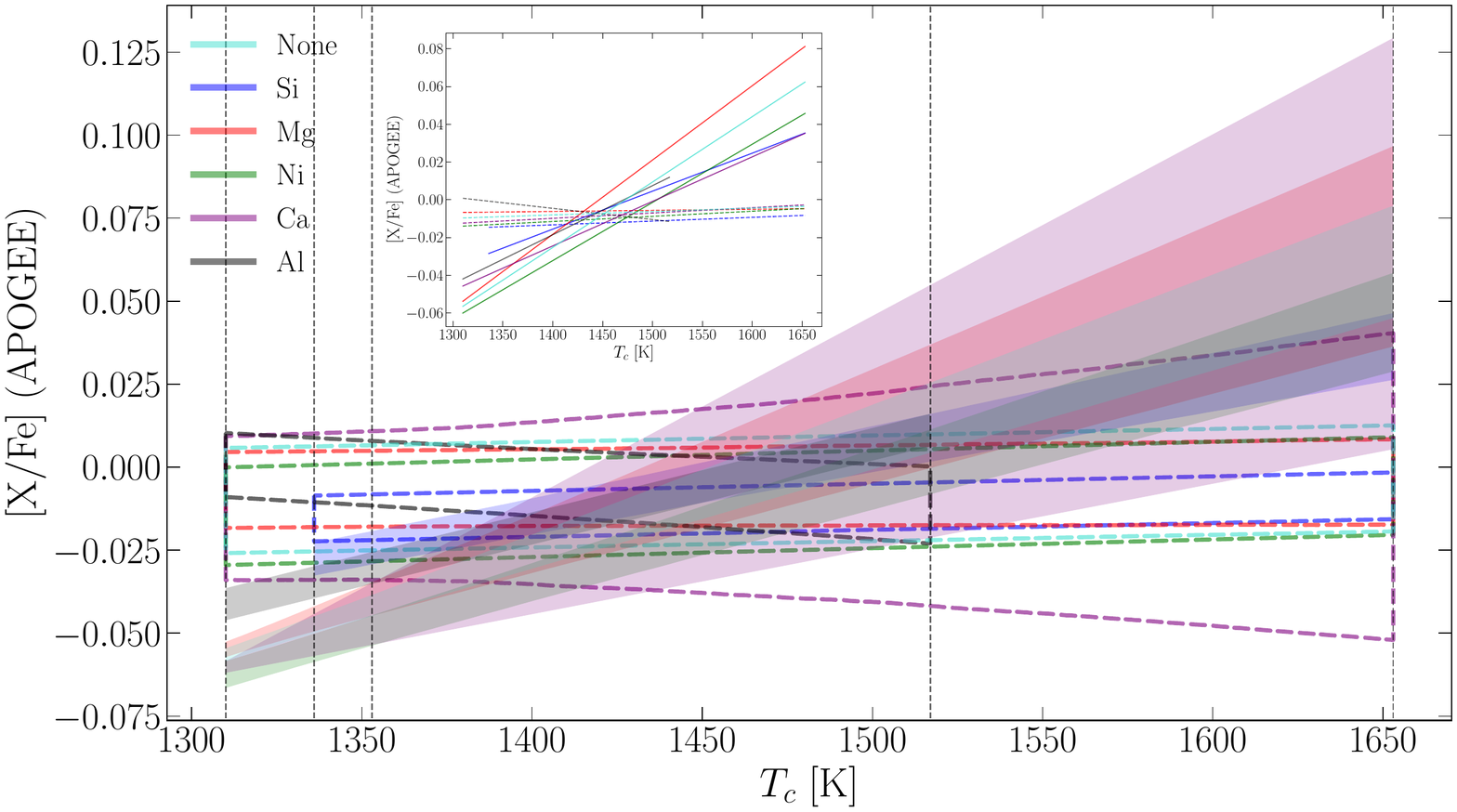}
    \centering
    \caption{Results on APOGEE data for the element exclusion test described in Appendix~\ref{sec: element exclusion test}. The legend specifies the excluded element, with the case of ``None" corresponding to no elements removed (i.e. Si, Mg, Ni, Ca, Al all included). Solid and dashed bands correspond to the ND and D model components, respectively. The inset plot illustrates the best fit mean abundance trends for both model components, with the form $\mu_m T_c + \mu_b$.}\label{fig: Leave One Out Test}
\end{figure*}

A more stringent test of the model constraints presented in \S\ref{sec: APOGEE Results} includes removing each element in turn, and evaluating the resulting fit across multiple trials. Constraints on $f$ for this test are presented in Fig.~\ref{fig: f constraint} and discussed in \S\ref{sec: APOGEE Results}. In this appendix, we provide an illustration of the depletion patterns for each trial, shown in Fig.~\ref{fig: Leave One Out Test}. Filled and unfilled bands correspond to not-depleted and depleted model components, respectively. The legend in Fig.~\ref{fig: Leave One Out Test} refers to the element removed, with ``None" representing the case for which all five elements are included (i.e. our main result in Fig.~\ref{fig:main result}). The inset plot illustrates the best fit mean abundance trend (e.g. $\mu_m T_c + \mu_b$).  

Constraints on the mean slopes and intercepts are robust, with $\sim 1\sigma$ shifts to be expected. The most significant differences stem from the standard deviation parameters $\sigma_m$ and $\sigma_b$, as can be seen by the varying band widths in Fig.~\ref{fig: Leave One Out Test}. Overall, however, the depletion patterns across all presented trials are similar, indicating that our analysis is robust to excluding a single element from our main results in \S\ref{sec: APOGEE Results}.

\section{Posterior on Model Parameters}\label{app:posterior on model parameters}
\begin{figure*}[ht!]
    \includegraphics[width=18.5cm]{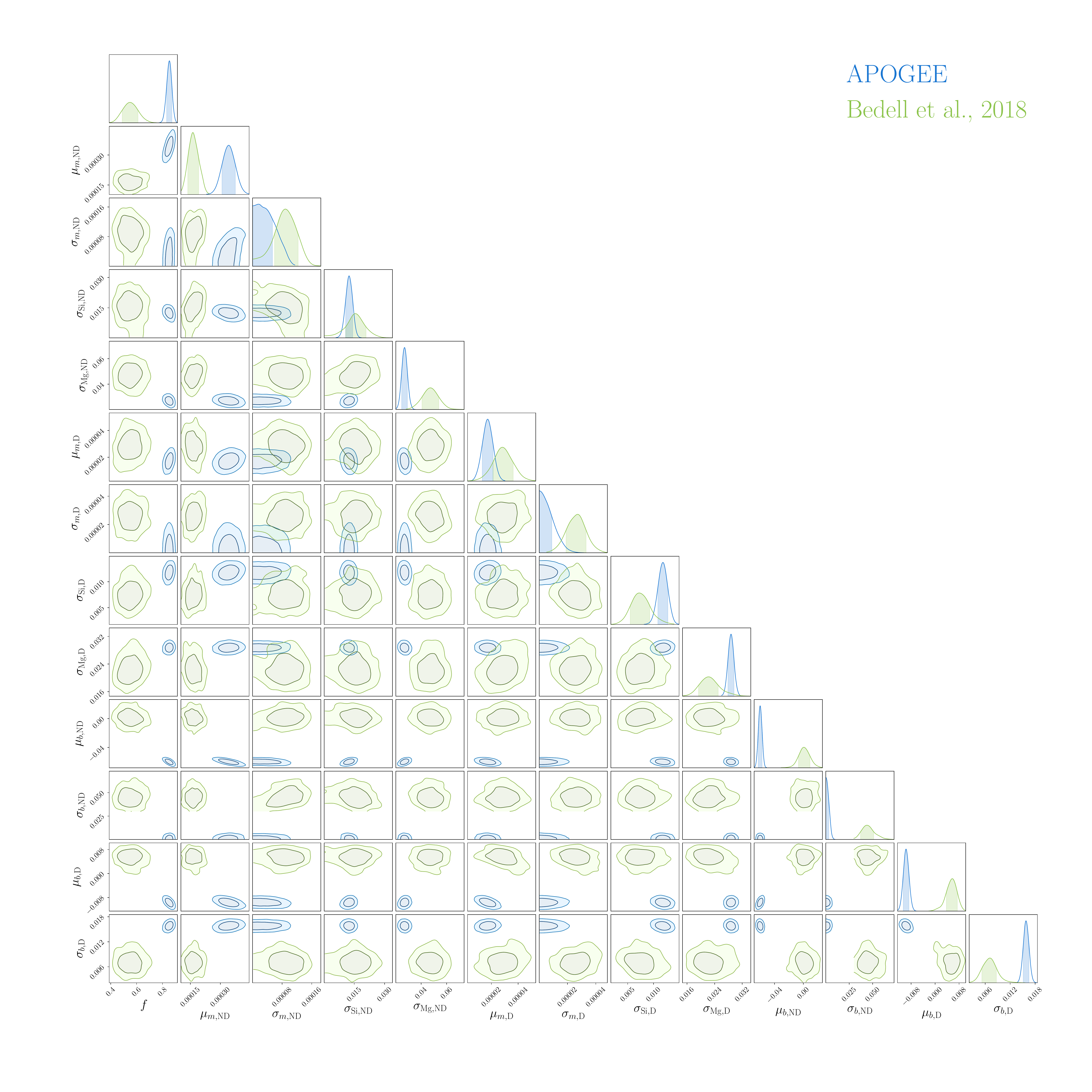}
    \centering
    \caption{Posterior on model parameters derived from APOGEE data (blue) and data from  \citeauthor{2018ApJ...865...68B} (green). Given the large number of parameters, we plot intrinsic scatter constraints for only Si and Al in each case. We note that the mean intercept parameter, $\mu_b$, is defined at $T_c = 1300~
    \rm{K}$ rather than zero in order to reduce degeneracy with the mean slope parameter, $\mu_m$.}\label{fig: Contours}
\end{figure*}

The posteriors on model parameters are illustrated in Fig.~\ref{fig: Contours} for results on data from APOGEE and \citet{2018ApJ...865...68B} in blue and green, respectively. Model parameters are subscripted with D and ND for depleted and not-depleted stars, respectively. Constraints on $\sigma_{\rm{intr},j}$ for Si and Mg are included in Fig.~\ref{fig: Contours} (labeled $\sigma_{\rm Si,\{D,ND\}}, \ \sigma_{\rm MG,\{D,ND\}}$), though we omit the inclusion of the other $\sigma_{{\rm intr},j}$ constraints since these parameters are not particularly meaningful to our present analysis.

We remind readers that some offsets in slope and intercept are to be expected, since APOGEE calibrates to solar neighborhood stars whereas \citet{2018ApJ...865...68B} calibrates directly to the Sun (see \S\ref{sec: Data APOGEE} and \S\ref{sec: caveats} for details). Furthermore, our APOGEE samples includes roughly 1700 stars with five elemental abundances for each, while the abundance data from \citet{2018ApJ...865...68B} contains many more elements with far fewer stars. We note that when limiting the abundance data from \citet{2018ApJ...865...68B} to the five overlapping elements used in our analysis of APOGEE data (Si, Mg, Ni, Ca, Al), the resulting constraints have a higher degree of overlap. We postpone the inclusion of more high $T_c$ elements in our APOGEE sample to a future work, since this analysis already relies upon high precision chemical abundances based on detailed discussions in  \citet{2020arXiv200705537J}.

\clearpage
\bibliography{thebib}
\end{document}